\documentclass[aps,prb,showpacs,citeautoscript,superscriptaddress,twocolumn]{revtex4-1}

\usepackage{graphicx}
\usepackage{amsmath}
\usepackage{amssymb}
\usepackage{color}

\newcommand{\bea}{\begin{eqnarray*}}
\newcommand{\eea}{\end{eqnarray*}}
\newcommand{\bne}{\begin{equation*}}
\newcommand{\ede}{\end{equation*}}

\newcommand{\bnen}{\begin{equation}}
\newcommand{\eden}{\end{equation}}
\newcommand{\bean}{\begin{eqnarray}}
\newcommand{\eean}{\end{eqnarray}}
\newcommand{\bnsn}{\begin{subequations}}
\newcommand{\edsn}{\end{subequations}}

\newcommand{\bna}{\begin{array}}
\newcommand{\eda}{\end{array}}
\newcommand{\bnm}{\begin{enumerate}}
\newcommand{\edm}{\end{enumerate}}
\newcommand{\bni}{\begin{itemize}}
\newcommand{\edi}{\end{itemize}}

\renewcommand{\vec}[1]{\text{\boldmath{$ #1 $}}}

\newcommand{\beff}{\mathcal{B}}
\newcommand{\vbeff}{\vec{\mathcal{B}}}

\newcommand{\ket}[1]{| #1 \rangle}
\newcommand{\bra}[1]{\langle #1 |}

\begin{document}
\title{Current hot spot in the spin-valley blockade in 
carbon nanotubes}

\author{G\'abor Sz\'echenyi}
\affiliation{Institute of Physics, E\"otv\"os University, Budapest, Hungary}

\author{Andr\'as P\'alyi}
\affiliation{Institute of Physics, E\"otv\"os University, Budapest, Hungary}
\affiliation{MTA-BME Exotic Quantum Phases Research Group,
Budapest University of Technology and Economics, Budapest, Hungary}

\date{\today}

\newcommand{\Dso}{\Delta_{\rm SO}}
\begin{abstract}
We present a theoretical study of the spin-valley blockade
transport effect in a double quantum dot defined 
in a straight carbon nanotube. 
We find that intervalley scattering due to short-range impurities
completely lifts the spin-valley blockade and induces a large
leakage current in a certain confined range of the external
magnetic field vector.
This current hot spot emerges due to
different effective magnetic fields acting on the spin-valley qubit 
states of the two quantum dots.
Our predictions are compared to a recent measurement
[F. Pei {\it et al.}, Nat. Nanotech. {\bf 7}, 630 (2012)].
We discuss the implications for blockade-based 
schemes for qubit initialization/readout, and
 motion sensing of nanotube-based
mechanical resonators. 
\end{abstract}


\maketitle


\section{Introduction}

Breakthrough experiments in the past decade have demonstrated
the ability to initialize, manipulate, couple and read out
spin-based quantum bits\cite{Loss-divincenzo} (qubits) using electrons in
electrostatically defined quantum dots (QDs)\cite{Elzerman-nature, 
Petta-science, Koppens-esr, Nowack-esr, 
Hanson-rmp}. 
A key ingredient in many of those experiments 
is the Pauli blockade
 mechanism\cite{Ono-spinblockade}.
 Pauli blockade is a characteristic feature of electronic transport through
 a double quantum dot (DQD) via the (1,1)$\to$(0,2)$\to$(0,1)$\to$(1,1)
 cycle of charge configurations, where ($n$,$m$) stands for 
 states with  $n$ electrons in the first QD and
 $m$   electrons in the second QD.
If a spin-triplet state is occupied in the (1,1) charge configuration,
then Pauli's exclusion principle prevents the (1,1)$\to$(0,2) tunneling
process
and thereby blocks the current flow. 
This simple mechanism allows for
initialization and readout of spin states via current or charge sensing 
measurements in a serially coupled double quantum dot (DQD). 
Pauli blockade measurements have also been utilized to experimentally
identify the strengths of spin-orbit and hyperfine interactions in
DQDs\cite{Koppens-spinblockade,NadjPerge-disentangling}.
By combining 
a DQD with a mechanical resonator, the Pauli
blockade mechanism can be exploited to convert the fast motional 
oscillations ($\sim 100$ MHz) of the resonator to
a direct current  through the DQD, enabling a simple
dc electronic  detection of the 
resonator's motion\cite{Ohm}.

\begin{figure}[h!]
\includegraphics[width=0.8\columnwidth]{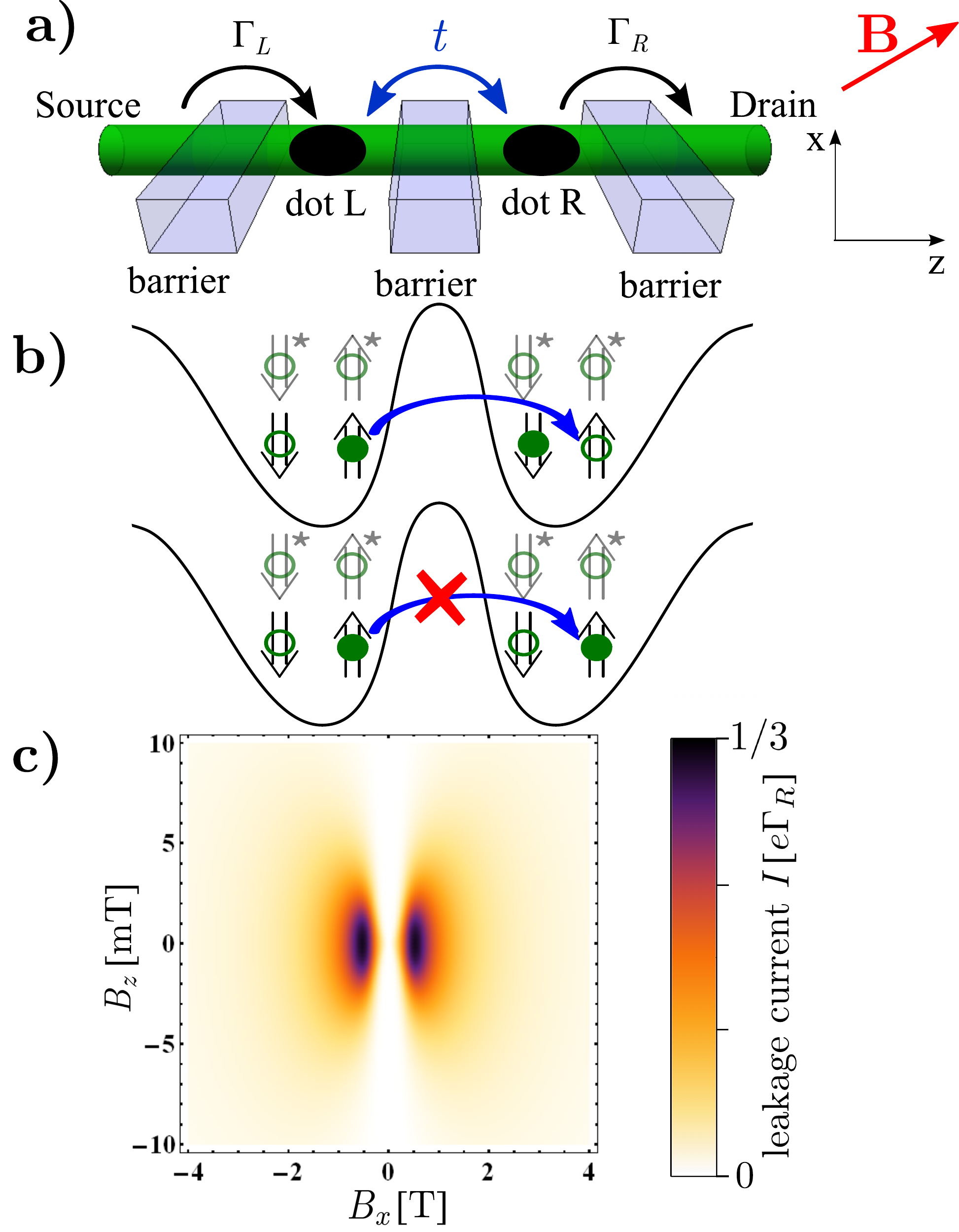}
\caption{\label{fig:Setup}
(Color online)
(a) Schematic  of the carbon nanotube double quantum dot
transport setup showing spin-valley blockade.  
The red arrow represents the external magnetic field $\vec B = (B_x,0,B_z)$. 
Lead-dot tunneling rates $\Gamma_L$, $\Gamma_R$ and the coherent
interdot tunneling
amplitude $t$ are indicated.
(b) Schematic of the energy levels involved in the transport cycle
$(1,1) \to (0,2) \to (0,1)$.
If two electrons form a triplet in the (1,1) charge configuration, 
then current becomes blocked due to Pauli's exclusion principle.
(c) Current hot spots at finite transverse magnetic field,
formed due to the complete lifting of the spin-valley blockade by 
short-range disorder. 
}
\end{figure}

Among the numerous host materials for quantum dots, 
carbon nanotubes\footnote{Throughout this work, we focus on 
\emph{semiconducting} nanotubes, as metallic tubes are
not suitable for hosting electrostatically defined quantum dots because of the
absence of an energy gap.} (CNTs) are unique because of the
simultaneous presence of the valley degree of freedom
of their electrons 
and the strong spin-orbit 
interaction\cite{Kuemmeth,Jespersen,Steele-strongsoi}.
The two-valued valley degree of freedom is related to the clockwise or
anti-clockwise circulating motion of the electron along the CNT 
circumference,
and is responsible for nominally fourfold degenerate (spin and valley)
orbital energy levels in electrostatically defined QDs  
(see Fig. \ref{fig:Setup}a and b).
The two valley states are typically denoted by $K$ and $K'$.
The main effect of the strong spin-orbit interaction is that it induces a large 
energy splitting $\Delta_{\rm so} \sim 0.1-3$ meV
within each fourfold degenerate orbital QD level.
At zero magnetic field the low-energy doublet,
depicted as $\Uparrow$  and $\Downarrow$  in Fig. \ref{fig:Setup}b, 
is formed by a time-reversed pair of states.
In the absence of valley mixing, 
$\Uparrow$ is an up-spin state circulating in one direction 
along the CNT circumference, and $\Downarrow$ is
a down-spin state circulating in the other direction. 

It is natural to think of the low-energy doublet 
$\Uparrow,\Downarrow$ as 
a \emph{spin-valley qubit}\cite{FlensbergMarcus,Laird}.
A resonant manipulation scheme for this qubit
 in a bent CNT has been proposed\cite{FlensbergMarcus}
and experimentally implemented\cite{Laird} recently.
Here again, the Pauli blockade mechanism, named 
\emph{spin-valley blockade}\cite{FeiPei,Laird,Palyi-hyperfine,Palyi-cnt-spinblockade} in this context, was used for qubit initialization and
readout.

Motivated by recent measurements in CNT DQDs\cite{Buitelaar,ChurchillPRL,ChurchillNPhys,Chorley,FeiPei,Laird},
and the potential experimental applications,
here we theoretically describe the spin-valley blockade
transport effect in a straight nanotube. 
The schematic view of such a CNT DQD device and the
blocking mechanism are shown in
Fig. \ref{fig:Setup}a and b, respectively.
Our quantity of interest is the direct current $I$, 
also known as the \emph{leakage current}, that flows
from the source to the drain 
through the DQD that is tuned to the spin-valley blockade regime.
We calculate the current $I$
as a function of the magnitude and direction of the
external magnetic field $\vec B$.
In our model we include spin-orbit interaction and short-range disorder,
 allow for both longitudinal and transverse 
vector components of the magnetic field
with respect to the CNT axis,
and use the two-site Hubbard model to 
describe interdot tunneling and 
the Coulomb repulsion between electrons on the DQD.
We focus on the case of clean devices, defined by the condition 
that the characteristic 
energy scale of short-range disorder
is exceeded by that of the spin-orbit interaction.

Our main result is that for a generic distribution of short-range
impurities, 
a \emph{current hot spot}, i.e., a region of high current,
appears
if the magnetic field vector is 
approximately transverse to the CNT axis, and its magnitude 
is tuned within a certain range.
An example is shown in Fig. \ref{fig:Setup}c,
where the current hot spots are located in the vicinity of 
$|B_x| \approx 0.5$ T.
The current hot spot emerges because  
the spin-valley blockade is completely lifted
due to the interplay of the short-range impurities and
the appropriately tuned transversal magnetic field.
Below we show that the transverse magnetic field corresponding to the 
center of the hot spot is proportional to the energy scales
of spin-orbit coupling $\Delta_{\rm so}$ and interdot tunneling $t$, and 
inversely proportional to the energy scale $\Delta_{KK'}$
of  short-range disorder [see Eq. \eqref{eq:bxhotspot}].
The current hot spot is most pronounced for zero 
energy detuning $\epsilon = 0$ between the (1,1) and (0,2) states,
and gradually disappears as the magnitude of detuning is increased
above the energy scale of the interdot tunneling. 
By utilizing the pseudospin-1/2 description of the spin-valley qubit
introduced by Flensberg and Marcus\cite{FlensbergMarcus}, 
and the master-equation
model of Pauli blockade in spinful DQDs 
developed in Refs.\, \onlinecite{Jouravlev} and \, \onlinecite{Coish},
we describe the blockade-lifting
mechanism both on a quantitative and a qualitative level. 
The mechanism found here is relevant for applications relying
on the Pauli blockade effect such as 
qubit initialization/readout\cite{Laird} and the dc electronic 
motion sensing of a CNT mechanical resonator\cite{Ohm} via
the qubit-phonon coupling\cite{RudnerRashba,Palyi-spinphonon}.

We note that our present work extends 
Ref. \onlinecite{Palyi-cnt-spinblockade} where the leakage current
was calculated in a longitudinal magnetic field. 
A number of further theoretical works studied distinct
characteristics of Pauli blockade in CNTs, including
descriptions of the pulsed-gated DQD experiments of 
Ref.~\onlinecite{ChurchillPRL}\cite{Reynoso1,Reynoso2},
the spectrum of two-electron single\cite{Wunsch,SecchiRontani} and double\cite{Weiss} QDs,
and the leakage current influenced by the 
formation of an electronic Wigner molecule\cite{vonStecher} and
by hyperfine interaction\cite{Palyi-hyperfine,Kiss}.

The rest of the paper is organized as follows. 
In Sec.~\ref{sec:effectivefield}, we reformulate the pseudospin-1/2
description\cite{FlensbergMarcus} of the single-electron spin-valley qubit in 
a single CNT QD. 
In Sec.~\ref{sec:model} we revisit 
the master-equation model\cite{Jouravlev,Coish}
 of the Pauli blockade,
and derive our central analytical formula for the leakage current. 
In Sec.~\ref{sec:results} we present and interpret our results,
which is followed by a discussion in 
Sec.~\ref{sec:discussion}.



\section{Effective magnetic field felt by the spin-valley qubit}
\label{sec:effectivefield}

Here we consider a single QD with a single electron occupying
the nominally fourfold degenerate (spin and valley) ground state
of an electrostatically defined CNT QD. 
Following Ref.\, \onlinecite{FlensbergMarcus}, 
we derive the effective magnetic 
field acting on the spin-valley qubit formed by the lower-lying
time-reversed pair of the four states.
The effective magnetic field arises as a combined effect of the
external magnetic field and disorder-induced valley mixing. 
The transport theory yielding the leakage current will be based on the
concept of the effective magnetic field in the subsequent Section.

The relative orientation of the CNT and the reference frame 
is shown in Fig.~\ref{fig:Setup}a.
The $4\times 4$ Hamiltonian describing the effects of
spin-orbit interaction, valley mixing, and external magnetic field
on a single spin-valley-degenerate QD level is 
$H=H_0+H_1$, where
\bnen
H_0 = -\frac{\Delta_{\rm so}}{2} \tau_3 s_z
\eden
and 
\bean
\nonumber
H_1 &=& 
\frac 1 2
\mathrm{Re}\left(\Delta_{KK'}\right) \tau_1 +
\frac 1 2 \mathrm{Im}\left(\Delta_{KK'}\right) \tau_2\\
&+& \frac 1 2 g_s \mu_B \vec B \cdot \vec s
+ \frac 1 2 g_v \mu_B B_z \tau_3.
\eean
Here $\Delta_{KK'} = |\Delta_{KK'}| e^{i\varphi}$ 
is the complex valley-mixing matrix 
element\cite{Palyi-cnt-spinblockade,Palyi-valley-resonance}
e.g., induced by short-range disorder, $\tau_1$, $\tau_2$ and $\tau_3$ 
($s_x$, $s_y$ and $s_z$)
are Pauli matrices acting in valley (spin) space,
$g_s \approx 2$ is the spin g-factor, $\mu_B$ is the Bohr-magneton, and
$\vec B = (B_x,0,B_z)$ is the external magnetic field.
Finally, $g_v$ is the valley g-factor, whose value depends on 
the chirality of the CNT and ranges approximately between 10 and 50
in experiments using clean CNT QDs
\cite{Minot,Kuemmeth,ChurchillPRL,Jespersen-orbital,FeiPei,Steele-strongsoi}.

Throughout this work we focus on the spin-orbit-dominated
regime of energy scales, i.e., 
 \bnen
 \label{eq:hierarchy}
\Delta_{\rm so} \gg 
\Delta_{KK'}, \
g_v \mu_B B_z, \ g_s \mu_B B_x.
\eden
(Comparisons of order-of-magnitudes, 
such as Eq. \eqref{eq:hierarchy}, correspond to the absolute values
of the involved quantities.)
This regime was achieved in recent 
experiments
using relatively clean 
CNTs\cite{Kuemmeth,ChurchillPRL,ChurchillNPhys,FeiPei,Laird}
showing weak valley mixing.
Assuming Eq. \eqref{eq:hierarchy}, we treat
$H_1$ perturbatively. 
The two-dimensional ground-state (excited-state) 
subspace of $H_0$ is
formed by the time-reversed pair 
$\ket{K \uparrow}$ and $\ket{K' \downarrow}$ 
($\ket{K\downarrow}$ and $\ket{K'\uparrow}$), 
with energy eigenvalue $-\Delta_{\rm so}/2$ ($\Delta_{\rm so}/2$).
In general, valley-mixing and the external magnetic field couples
the ground-state and excited-state subspaces. 
Due to Eq.~\eqref{eq:hierarchy}, the coupling between the
ground-state and excited-state subspaces 
can be eliminated by an appropriately
chosen (Schrieffer-Wolff) unitary transformation\cite{Winkler}
of the four-dimensional
Hilbert space.
This transformation results a $2\times 2$ effective Hamiltonian
$H_{\rm eff}$
describing the dynamics within the perturbed ground-state subspace,
allowing to describe the electron in that subspace as a
spin-1/2 particle in an effective magnetic (Zeeman) field.  

The effective Hamiltonian of the ground-state subspace
is obtained
via the second-order Schrieffer-Wolff transformation\cite{Winkler}
$U_{\rm SW} = e^{-S}$ , 
with
\bnen
S=\frac{1}{2\Delta_{SO}}\left[\begin{matrix} 0 & 0 & -g_s\mu_BB_x & -\Delta_{KK'}\\ 0 & 0 & -\Delta_{KK'}^*&-g_s\mu_BB_x\\ g_s\mu_BB_x&\Delta_{KK'}&0&0\\\Delta_{KK'}^*& g_s\mu_BB_x&0&0 \end{matrix}\right],
\eden
where the basis $\left(\ket{K' \uparrow},\ket{K \downarrow},\ket{K' \downarrow},\ket{K\uparrow}\right)$ is used.
This transformation approximately 
decouples the ground-state and excited-state
subspaces, resulting in the following effective Hamiltonian for the
ground-state subspace:
\bnen
\label{eq:effH}
H_{\rm eff} = 
\beff_1 \sigma_1 + \beff_2 \sigma_2
+ \beff_3 \sigma_3
\equiv \vbeff \cdot \vec{\sigma},
\eden
where 
\begin{subequations}
\label{eq:beffcomponents}
\bean
\beff_1 &=&
\frac{g_s \mu_B B_x |\Delta_{KK'}| \cos\varphi}{2\Delta_{\rm so}}, \\
\beff_2 &=&
\frac{g_s \mu_B B_x |\Delta_{KK'}| \sin\varphi}{2\Delta_{\rm so}}, \\
\beff_3 &=& 
\frac 1 2 \left(g_v + g_s \right) \mu_B B_z,
\eean
\end{subequations}
and $\sigma_i$ is the $i$-th Pauli matrix acting in the
perturbed two-dimensional subspace spanned by
\begin{subequations}
\label{eq:pseudospin}
\bean
\ket{\Uparrow} &=&
\ket{K \uparrow} 
- 
\frac{g_s \mu_B B_x }{2\Delta_{\rm so}} \ket{K\downarrow}
+
\frac{\Delta_{KK'} }{2\Delta_{\rm so}} \ket{K' \uparrow} ,
\\
\ket{\Downarrow} &=& 
\ket{K' \downarrow} 
+ \frac{\Delta_{KK'}^*}{2\Delta_{\rm so}} \ket{K \downarrow}
- 
\frac{g_s \mu_B B_x }{2\Delta_{\rm so}} \ket{K'\uparrow}.
\eean
\end{subequations}
Furthermore, $\vec{\sigma} = (\sigma_1,\sigma_2,\sigma_3)$, and 
$ \vbeff  = (
\beff_1,\beff_2,\beff_3)$.

Naturally, the effective Hamiltonian $H_{\rm eff}$ in Eq. \eqref{eq:effH}
takes the form of a Zeeman Hamiltonian
describing a spin-1/2 particle in a magnetic field.
Accordingly, we will refer to the two basis states of 
Eq.~\eqref{eq:pseudospin} as representing
a \emph{pseudospin}. 
For brevity, the effective magnetic field $\vbeff$ is defined
in energy units.
Note that the first two components of the effective magnetic field
$\vbeff$ are nonzero only if both the valley mixing and the
transverse magnetic field are nonzero. 
Furthermore, because of the perturbative character of 
the first two components of $\vbeff$, 
the effective Hamiltonian is dominated by $\beff_3$
unless the external B-field is directed almost perfectly or perfectly
along the transversal-to-CNT direction.

In contrast to Ref. \, \onlinecite{FlensbergMarcus},
here we kept track of the phase $\varphi$ of the complex valley-mixing
matrix element $\Delta_{KK'}$, which influences the first two components
of the effective magnetic field $\vbeff$. 
This phase $\varphi$ has no physical significance in a single QD, 
since its value changes upon multiplying one of the low-energy basis
states with an arbitrary complex phase factor.
Nevertheless, the difference of the $\varphi$ phases in two QDs $L$ and $R$, 
i.e., $\Delta \varphi = \varphi_L - \varphi_R$, does have physical significance. 
For example, this phase difference influences 
 the leakage current in spin-valley blockade, as shown in Fig.~\ref{fig:lepedo}.
(For further examples, see, eg,
Refs.~\onlinecite{Palyi-cnt-spinblockade,
Palyi-valley-resonance,Reynoso2,CulcerPRL,YueWu})

\section{Leakage current in spin-valley blockade}
\label{sec:model}

In this Section, we rely on the notion of effective magnetic field 
$\vbeff$ to calculate the leakage
current through a CNT DQD under spin-valley blockade.
To this end, we specify the transport problem,
and utilize the model introduced in Ref.\, \onlinecite{Jouravlev},
and the classical master equation outlined in Ref.\, \onlinecite{Coish}, 
to derive an analytical result for the leakage current.
Conlusions are drawn, and comparison is made
to experimental data, in Section \ref{sec:results}. 

Importantly, we consider the case when only the lower-lying
time-reversed pairs of each dot of the DQD participate in transport, 
i.e., the states $\Uparrow^*$ and $\Downarrow^*$ in 
Fig. \ref{fig:Setup}b are disregarded.
This case is realized if the source-drain
bias voltage and the DQD energy levels are tuned appropriately. 
In this case,
there are 7 states that participate in transport,
in complete analogy to spin blockade in GaAs\cite{Jouravlev}.
Two of them are single-electron states in the (0,1) charge configuration:
$\ket{0,\Uparrow}$ and $\ket{0,\Downarrow}$.
Four of them are (1,1) states and there is a single (0,2) state 
$\ket{S_g} \equiv \ket{0,\Uparrow \Downarrow}$,
adding up to 5 two-electron states in total. 
For the (1,1) states, we will use both the product basis 
$\ket{\Uparrow,\Uparrow}$,
$\ket{\Uparrow,\Downarrow}$,
$\ket{\Downarrow,\Uparrow}$,
$\ket{\Downarrow,\Downarrow}$,
and the
singlet-triplet basis
\bean
\ket{S} &=& \frac 1 {\sqrt{2}} \left(\ket{\Uparrow,\Downarrow} - \ket{\Downarrow,\Uparrow}\right), \\
\ket{T_0} &=& \frac 1 {\sqrt{2}} \left(\ket{\Uparrow,\Downarrow} +\ket{\Downarrow,\Uparrow}\right), \\
\ket{T_+} &=& \ket{\Uparrow,\Uparrow}, \\
\ket{T_-} &=& \ket{\Downarrow,\Downarrow}.
\eean

The Hamiltonian describing the DQD is
\bnen
\label{eq:dqdhamiltonian}
H_{\rm DQD} = H_t + H_{\mathcal B} + H_\epsilon.
\eden
Here, $H_t$ represents tunneling between the two QDs.
We assume spin- and valley-conserving tunneling, which 
is represented by 
$H_t = \sqrt{2} t (\ket{S_g}\bra{S} + \ket{S}\bra{S_g})$,
with $t$ being the tunnel amplitude.
Strictly speaking, the spin- and valley-conserving property does not 
imply the conservation of the pseudospin. 
Nevertheless, the pseudospin-flip interdot tunneling amplitude
is much smaller than $t$, hence we disregard it.
The effective magnetic fields, induced by short-range disorder and
the external magnetic field, are incorporated in the second 
Hamiltonian term
\bnen
H_{\mathcal B} = 
	\vbeff_{L} \cdot \vec{\sigma_L} +
	\vbeff_{R} \cdot \vec{\sigma_R}.
\eden
Recall that the short-range disorder configuration on dot $L$ is
independent of that on dot $R$, and therefore
the disorder-related components 
[see Eq.~\eqref{eq:beffcomponents}] of $\vbeff_L$ 
are independent of those of $\vbeff_R$.
The term $H_\epsilon = \epsilon \ket{S_g}\bra{S_g}$ 
represents the gate-controlled 
energy detuning between the (1,1) and (0,2) charge configurations.
We focus on the zero-detuning case $\epsilon = 0$ in this Section,
and discuss the
case $\epsilon \neq 0$ in Sec. \ref{sec:results}.

Once the eigenstates of $H_{\rm DQD}$ are known,
the dynamics of current flow can be described by the classical 
master equation\cite{Coish}
\begin{subequations}
\label{eq:cme}
\bean
\dot{p}_\alpha &=& -  \left(\sum_j W_{j \leftarrow \alpha}\right) p_\alpha 
+  \sum_j  W_{\alpha \leftarrow j} p_j,\\
\dot{p}_j &=& - \left( \sum_\alpha W_{\alpha \leftarrow j} \right) p_j
+  \sum_\alpha  W_{j \leftarrow \alpha} p_\alpha.
\eean
\end{subequations}
Here, index $\alpha \in \{1,2,\dots,5\}$ 
(index $j\in\{1,2\}$)
represents two-electron (single-electron) eigenstates of $H_{\rm DQD}$,
$p_{\alpha/j}$ are occupation probabilities summing up
to unity, i.e., $\sum_{\alpha=1}^5 p_\alpha + \sum_{j=1,2} p_j = 1$,
and $W_{\alpha \leftarrow j}$ ($W_{j \leftarrow \alpha}$) are
transition rates representing electron tunneling to the
DQD from the left contact (from the DQD to the right contact).

The transition rates are expressed from Fermi's Golden Rule
as
\begin{subequations}
\label{eq:fermigoldenrule}
\bean
W_{\alpha \leftarrow j} &=& \Gamma_L \sum_{\sigma=\Uparrow,\Downarrow} \left|\bra{\alpha}  d^\dag_{L\sigma} \ket{j} \right|^2,\\
W_{j \leftarrow \alpha} &=& \Gamma_R \sum_{\sigma=\Uparrow,\Downarrow} \left| \bra{j} d_{R\sigma} \ket{\alpha} \right|^2,
\eean
\end{subequations}
where, e.g., $d_{L\Uparrow}$ is an electron operator creating 
an electron on dot $L$ with pseudospin $\Uparrow$.
The rate $\Gamma_L$ ($\Gamma_R$) is the single-electron
tunneling rate at the left (right) contact.
The leakage current in the steady state is given by 
\bnen
\label{eq:current}
I = \sum_{\alpha j} W_{\alpha \leftarrow j} \bar p_j,
\eden
where $\bar p_j$ is the steady-state occupation probability of
the single-electron state $j$.

We are able to analytically diagonalize $H_{\rm DQD}$,
and therefore to obtain an analytical formula for the leakage current.
The result is expressed with the symmetric and antisymmetric
combinations of the effective magnetic fields,
\bnen
\vbeff_s = \frac 1 2 \left(
	\vbeff_L	+ \vbeff_R
\right)
\eden
and
\bnen
\vbeff_a = \frac 1 2 \left(
	\vbeff_L - \vbeff_R
\right),
\eden
respectively.
The resulting formula for the leakage current is
\begin{subequations}
\label{eq:nonpert}
\begin{eqnarray} 
\frac{I}{e\Gamma_R}  &=&
\left[
	\frac{t^2}{4 \beff_a^{\| 2}}+\frac{F(\beff_s,\vbeff_a)}{4t^2\beff_a^{\perp 2}}-
	\frac{1}{2} + \frac{\Gamma_R}{2\Gamma_L}, \right]^{-1}\\
	F(\beff_s,\vbeff_a) &=& (\beff_s^2+\beff_a^2+2t^2)^2-
	4\beff_s^2(2t^2+\beff_a^{\parallel 2}).
\end{eqnarray}
\end{subequations}
Here, the vector $\vbeff_a^\parallel$ (the vector $\vbeff_a^\perp$) 
is the projection of 
$\vbeff_a$ 
onto the direction of $\vbeff_s$, 
(orthogonal to $\vbeff_s$).

Note that our analytical result \eqref{eq:nonpert} is valid 
irrespective of the energy scale hierarchy between 
$\mathcal B_a$, $\mathcal B_s$ and $t$. 
In this sense, Eq. \eqref{eq:nonpert} interpolates between
the zero-detuning limits of the perturbative
results Eq. (6) of Ref. \onlinecite{Jouravlev}
and Eq. (8) of Ref. \onlinecite{Jouravlev},
the former (latter) being valid if
$\mathcal B_a \ll t,\mathcal B_s$,
($\mathcal B_a,\mathcal B_s \gg t$).
Equation \eqref{eq:nonpert} also incorporates the dependence of
the leakage current on the tunneling rate $\Gamma_L$ at the
left lead-dot barrier.
In the special case  $\Gamma_L \gtrsim \Gamma_R$ and 
$B_a\ll t,B_s$, our Eq. \eqref{eq:nonpert} simplifies to 
\bnen
\label{eq:jn6}
\frac{I}{e\Gamma_R}=
\left[
	\frac{t^2}{4 \beff_a^{\parallel 2}}+\frac{\left(\beff_s^2-2t^2\right)^2}{4t^2\beff_a^{\perp 2}}\right]^{-1}.
\eden
Note that this formula is not identical to Eq. (6) of Ref. \onlinecite{Jouravlev}.
Difference in the magnitudes of constant factors probably arise from 
the different definitions of the parameters of the Hamiltonian.
In addition, a physically relevant difference is the minus sign in Eq. 
\eqref{eq:jn6},
which substitutes a corresponding plus sign of Eq. (6) of 
Ref. \onlinecite{Jouravlev}. 
Equation \eqref{eq:jn6}  suggests a resonant enhancement
of the leakage current at $|\beff_s| = \sqrt{2}t$.
Such an enhancement is 
indeed expected, since in this case
the triplet states polarized parallel or antiparallel to 
$\vbeff_s$ match the (1,1)-(0,2) hybrid singlet states in energy.
Hence we think that the minus sign in Eq. \eqref{eq:jn6} is correct.
For the weak-tunneling case $\beff_a,\beff_s \gg t $, Eq. \eqref{eq:nonpert}
implies 
\bnen 
\label{eq:jn8}
	\frac{I}{e\Gamma_R}=\frac{t^2}{\beff_s^2} \left(\vec n_L \times\vec n_R\right)^2,
\eden
where the vectors $\vec n_{L/R} = \frac{\vbeff_{L/R}}{\beff_{L/R}}$
are the unit vectors associated to the effective magnetic field
vectors in the two QDs.
Up to a constant of unit order of magnitude, this formula matches
the corresponding result Eq. (8) of Ref. \onlinecite{Jouravlev}.
Note that Eqs. \eqref{eq:nonpert}, \eqref{eq:jn6} and \eqref{eq:jn8}
were also verified by comparison to the corresponding numerical results. 

We note that the classical master equation \eqref{eq:cme} is
appropriate for describing the transport process only if the
energy distances between the eigenvalues of $H_{\rm DQD}$ exceed
the energy scales $h \Gamma_{L,R}$ 
associated to the lead-DQD tunnel rates. 
In certain cases, e.g., in the presence of level degeneracies,
it might be necessary to use a quantum master equation to
model the transport process. 
A particular example of Pauli blockade where spectral degeneracies
are important, and a quantum master equation  is needed, 
is treated in Ref. \onlinecite{Jeroen}.

\section{Results}
\label{sec:results}

\subsection{Current hot spot}

\begin{figure*}
\includegraphics[width=0.9\textwidth]{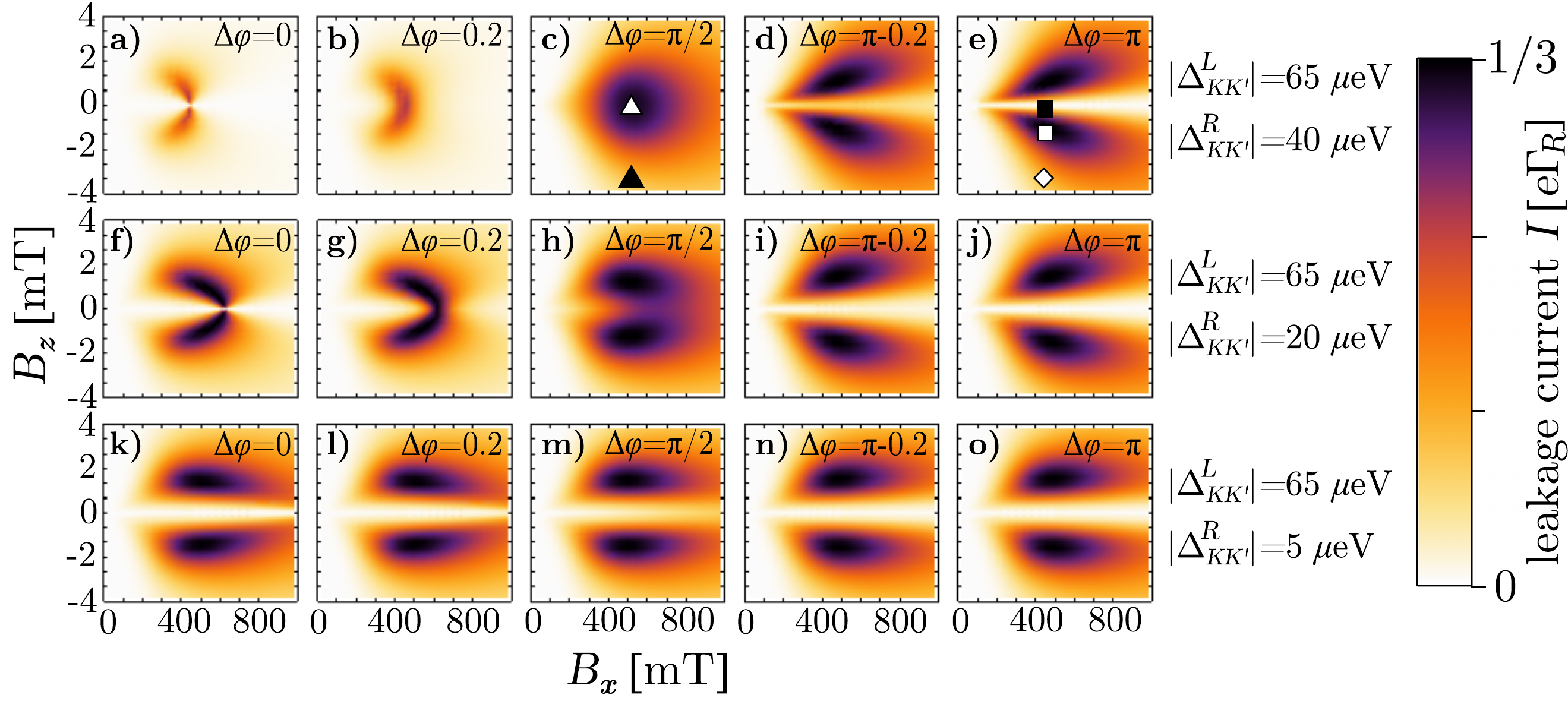}
\caption{\label{fig:lepedo}
(Color online)
Current hot spot for various values of the valley-mixing matrix elements
$\Delta_{KK'}^{L}$ and $\Delta_{KK'}^{R}$ in the two dots,
in the case of coherent interdot tunneling.
Each row of graphs is obtained using the valley-mixing
matrix element magnitudes given at the right end of the row. 
The difference $\Delta \varphi$ of the complex phases
of the valley-mixing matrix elements is shown in each graph
at the top right corner. 
Parameters: 
$g_s = 2$, $g_v = 54$, $\Delta_{\rm so} =$370$\;\mu$eV, $t=$5$\;\mu$eV,
$\Gamma_L = \Gamma_R$.
The plots are obtained by evaluating Eq. \eqref{eq:nonpert}.
The leakage current values at the marked points 
($\triangle$, $\blacktriangle$, $\blacksquare$, $\square$, $\diamond$) are 
discussed in the text. 
}
\end{figure*}

The leakage current as a function of the external magnetic field
is shown in Fig. \ref{fig:lepedo}a-o, for various values of the valley-mixing
matrix elements $\Delta_{KK'}^{L}$ and $\Delta_{KK'}^{R}$.
(From now on, we redefine $\Delta_{KK'}$ as
$\Delta_{KK'} := \max \{|\Delta_{KK'}^{L}|,|\Delta_{KK'}^{R}|\}$)
This figure is based on our analytical result Eq. \eqref{eq:nonpert}.
In all plots of Fig. \ref{fig:lepedo}, 
current hot spots (magnetic-field regions with strongly 
enhanced leakage current) develop. 
In all plots, the maximum of the leakage current approaches the
order of magnitude of $e \Gamma_R$, indicating 
that the spin-valley blockade is completely lifted in the area of the
hot spot. 
The shape of the hot spot varies with the values of the valley-mixing
matrix elements.
The presence of these current hot spots is 
the central result of this work. 

The existence of the current hot spots has a simple  interpretation,
allowing us to estimate
(i) the location of the hot spot along the $B_x$ axis,
(ii) the lateral extension of the hot spot along the $B_x$ and $B_z$ axes,
and
(iii) the upper bound of the leakage current.

\begin{figure}
\includegraphics[width=\columnwidth]{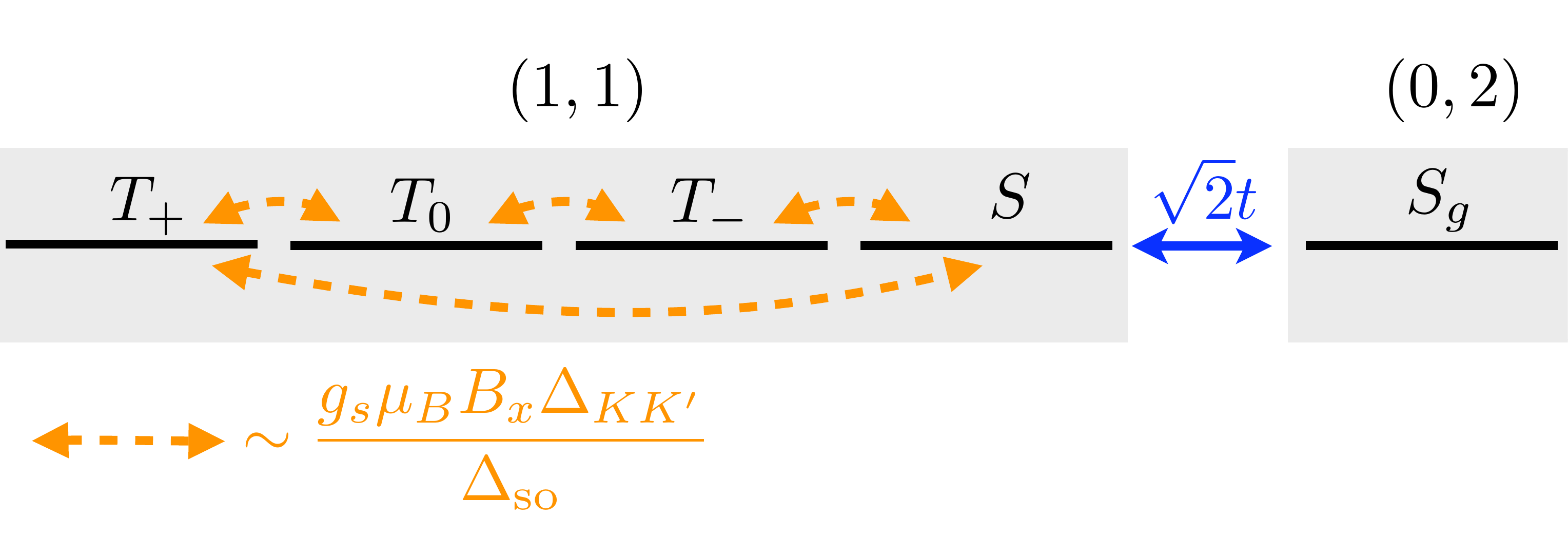}
\caption{
(Color online)
\label{fig:SimpleLevelScheme}
Schematic energy diagram of the two-electron
sector of the DQD Hamiltonian $H_{\rm DQD}$ at 
longitudinal magnetic field $B_z=0$ and zero detuning
$\epsilon = 0$.
The solid blue arrow represents the interdot tunneling $H_t$,
whereas the dashed orange lines represent the coupling matrix 
elements of $H_\mathcal{B}$ originating from the  effective magnetic fields
in the two dots. 
The latter matrix elements are typically of the same order of magnitude
$\sim \frac{g_s \mu_B B_x \Delta_{KK'}}{\Delta_{\rm so}}$.
}
\end{figure}

Consider the level scheme of the two-electron states shown in Fig. \ref{fig:SimpleLevelScheme},
which corresponds to the case of zero longitudinal magnetic field,  
$B_z =0$.
The horizontal lines of the level scheme represents the singlet-triplet
basis states: $\ket{T_+}$, $\ket{T_0}$, $\ket{T_-}$,
$\ket{S}$ and $\ket{S_g}$.
The arrows represent the Hamiltonian matrix elements that
couple these basis states.
At $B_x = 0$ and $t \neq 0$, the only coupling matrix element is
tunneling, denoted by the blue arrow.
By switching on $B_x$, the disorder-induced first and second 
components of the effective magnetic fields 
[see Eq. \eqref{eq:beffcomponents}] are switched on in both QDs.
Importantly, these effective magnetic fields appear in the 
singlet-triplet basis as off-diagonal Hamiltonian matrix elements 
mixing the triplets with the singlet $S$\cite{Jouravlev}.
The corresponding four matrix elements are depicted in 
Fig. \ref{fig:SimpleLevelScheme}
as dashed orange arrows.
These four matrix elements are usually unequal, but typically all of them
are of the same order of magnitude, 
$\sim \frac{g_s \mu_B B_x \Delta_{KK'}}{\Delta_{\rm so}}$.

Using the level structure in Fig. \ref{fig:SimpleLevelScheme},
we now argue that the leakage current is small, i.e., much smaller than
$e\Gamma_R$, if either
$\frac{g_s \mu_B B_x \Delta_{KK'}}{\Delta_{\rm so}} \ll t$
or
$\frac{g_s \mu_B B_x \Delta_{KK'}}{\Delta_{\rm so}} \gg t$.
In the former case, the (1,1) and (0,2) singlets $S$ and $S_g$ hybridize,
and the bonding (antibonding) state acquires a negative (positive) energy   
of the magnitude $\sqrt{2} t$. 
The singlet-triplet coupling matrix elements are much smaller than the
energies of the hybridized singlets, and therefore the coupling of the 
triplets to the singlets is only perturbative and hence very weak.
This implies that once any of the triplet states is occupied during transport, 
the flow of electrons is blocked for a long time, hence the 
time-averaged current is low. 
In the latter case, the spectrum becomes dominated by 
the effective magnetic fields on the two dots, 
the four energy eigenstates corresponding to the (1,1) sector
being $\pm \beff_L \pm \beff_R$.
The tunnel coupling to the (0,2) singlet $S_g$ is weak in this case,
implying a strongly suppressed leakage current.
This implies that the current hot spot is confined along the $B_x$
axis to the region where 
\bnen
\label{eq:bxhotspot}
B_x \sim \frac{t \Delta_{\rm so}}{g_s \mu_B \Delta_{KK'}}.
\eden

In all cases shown in Fig. \ref{fig:lepedo}, the switch-on of a sufficiently strong 
longitudinal magnetic-field component $B_z$
restores the spin-valley blockade.
The reason is that a strong $B_z$ energetically splits the polarized triplets
$\ket{T_+}$ and $\ket{T_-}$ from the singlets, making the hybridization of 
the former ones with the latter ones rather weak, and therefore
$\ket{T_+}$ and $\ket{T_-}$ will block the current flow.
This happens if $(g_v + g_s) \mu_B B_z \gg t$, hence the current
hot spot is confined along the $B_z$ axis to the range
\bnen
B_z \lesssim \frac{t}{(g_v + g_s) \mu_B}.
\eden

The upper bound of the leakage 
current for the case $\Gamma_L = \Gamma_R$
can be estimated as follows. 
It is plausible to assume, and possible to show formally, that the 
leakage current is maximal when each of the 5 two-electron energy
eigenstates has a $1/5$ weight in the (0,2) subspace. 
In this case, the decay rate of each two-electron state
is $2\Gamma_R/5$, whereas the decay rate of both one-electron
states is $2\Gamma_L=2\Gamma_R$.
Therefore the average time needed for a complete transport cycle
is $T = \frac{5}{2\Gamma_R} + \frac{1}{2\Gamma_R}$,
implying a leakage current of $I=e/T = \frac 1 3 e\Gamma_R$.

The shape of the current hot spot in Fig. \ref{fig:lepedo}
changes as the values of $\Delta_{KK'}^{L}$ and $\Delta_{KK'}^{R}$ are
changed;
e.g., in Fig. \ref{fig:lepedo}c, the hot spot has a circular shape, whereas
in Fig. \ref{fig:lepedo}e, current is low along the $B_x$ axis but
it is high in the two dark wing-shaped regions. 
Such variations of the current can be explained by 
analyzing the orders-of-magnitude of the
quantities appearing in Eq. \eqref{eq:nonpert}.
Here we focus on the five marked points of 
 Fig. \ref{fig:lepedo}c and e.
 
In case $\triangle$, the longitudinal external magnetic field $B_z$
is zero, hence the magnitudes and the enclosed angle of the 
effective magnetic fields $\vbeff_L$ and $\vbeff_R$ are
set by the relative magnitudes and complex phase angles of 
$\Delta_{KK'}^L$ and $\Delta_{KK'}^R$.
A straightforward evaluation of the parameters appearing
in Eq. \eqref{eq:nonpert} show that $t$, $B_s$, $B_a^{\parallel}$ and
$B_a^\perp$ all have the same order of magnitude, and therefore
the leakage current is of the order of $e\Gamma_R$. 
In case $\blacktriangle$, the longitudinal magnetic field $B_z$
is strong enough to dominate the effective magnetic fields.
Therefore, the antisymmetric combination of the effective magnetic fields
$\vbeff_a$ is almost perpendicular to the symmetric combination $\vbeff_a$,
implying $\beff_{a}^{\parallel} \ll \beff_a^{\perp}$, $\beff_s$, $t$.
This implies that the first term in the square bracket 
of Eq. \eqref{eq:nonpert} is much larger than unity, leading
to a leakage current in $\blacktriangle$ that is
 much smaller than $e\Gamma_R$.

In case $\blacksquare$, the longitudinal field is $B_z$ = 0.
This fact together with Eq. \eqref{eq:beffcomponents} imply that
the angle enclosed by the effective magnetic fields is the same 
as the relative complex phase $\Delta \varphi$ of 
the valley-mixing matrix elements, i.e., $\Delta \varphi = \pi$.
This implies that $\beff_a^{\perp} = 0$, which in turn implies
that the second term in the square bracket of Eq. \eqref{eq:nonpert}
diverges.
Therefore the current is zero at $\blacksquare$, even though
this point is at the center of the current
hot spot region.
In case $\square$, however, the finite $B_z$ tilts the effective 
magnetic fields and thereby reduces their enclosed angle,
rendering $t$ and the effective field components
on the rhs of Eq. \eqref{eq:nonpert}
comparable to each other.
Hence the current is large in $\square$.
Upon increasing $B_z$ further to point $\diamond$, the
enclosed angle of $\vbeff_L$ and $\vbeff_R$ approaches zero,
hence the current is suppressed for the same reason
as in case $\blacktriangle$.
Similar considerations can be used for the other subplots
of Fig. \ref{fig:lepedo} to interpret 
the current variations within the hot spot region.

\subsection{Detuning-dependence of the leakage current}
\label{subsec:detuning}

Our key analytical result Eq. \eqref{eq:nonpert} as well as our
Fig. \ref{fig:lepedo} are valid if the
energy detuning $\epsilon$ between the 
(1,1) states and the (0,2) singlet state $S_g$ is zero
(at zero B-field and zero interdot tunneling), i.e., 
if these states are aligned in energy.
However, this energy detuning is one of the easily tunable parameters 
in an experiment\cite{FeiPei}, hence it is desirable to know how the current
hot spot changes as the detuning $\epsilon$ is tuned away from zero. 

First we provide a brief, qualitative discussion.
The detuning $\epsilon$ is built into the DQD Hamiltonian Eq. 
\eqref{eq:dqdhamiltonian} as 
$H_\epsilon = \epsilon \ket{S_g}\bra{S_g}$.
At $\epsilon=0$, in the current hot spot region,
the condition $(g_v+g_s) \mu_B B_z \lesssim \frac{g_s \mu_B B_x \Delta_{KK'}}{\Delta_{\rm so}}\sim t$ guarantees the efficient
mixing of the 5 two-electron states, which in turn 
renders the leakage current large. 
This fact is unchanged by the
switch-on of $\epsilon$, as long as the 
order of magnitude of the latter 
does not exceed that of $t$.
If, however, $t \ll \epsilon$, then the hybridization of (1,1) states and
$S_g$ becomes only perturbative ($\sim t/\epsilon \ll 1$), 
and therefore the current hot spot disappears 
for such a strong detuning. 

\begin{figure}
\includegraphics[width=0.9\columnwidth]{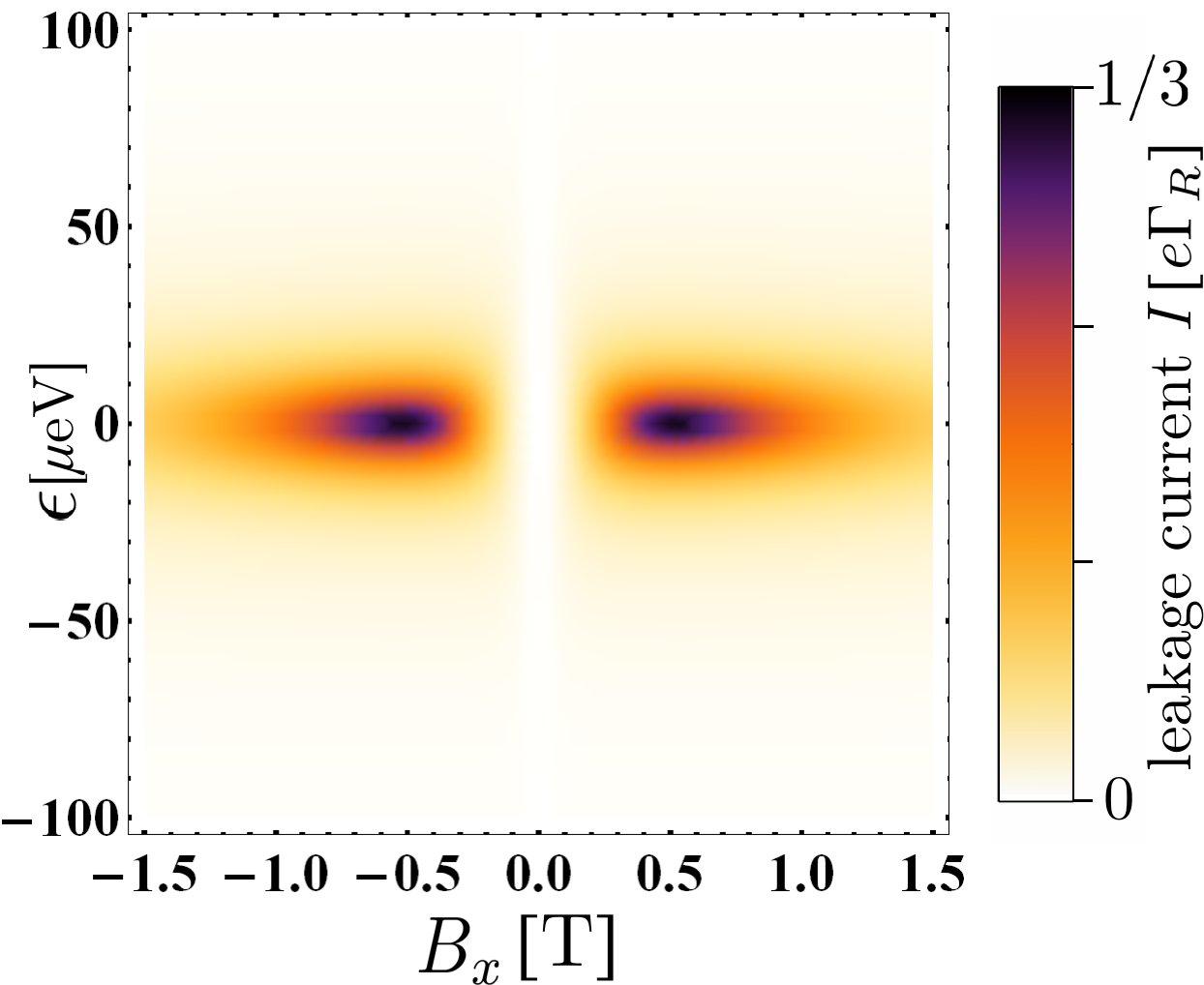}
\caption{\label{fig:detuning}
(Color online)
Leakage current as a function of transverse
external magnetic field and (1,1)-(0,2) energy detuning.
Parameters: $B_z = 0$; further parameter values are the same as 
for Fig. \ref{fig:lepedo}c.
}
\end{figure}

This behavior is shown in Fig. \ref{fig:detuning}. 
The plot is generated using Eq. \eqref{eq:current},
with transition rates calculated from the numerically obtained
eigenstates of $H_{\rm DQD}$ defined in Eq. \eqref{eq:dqdhamiltonian}.
The leakage current shown in Fig. \ref{fig:detuning}
displays the hot-spot feature in its dependence on $B_x$, 
and the decreasing current for $\epsilon \gg t$ as predicted
in the preceding paragraph. 

Figure \ref{fig:detuning} can be compared to the 
experimental data of Ref. \onlinecite{FeiPei},
where spin-valley blockade was observed 
and the magnetic field dependence of the leakage current was
studied in detail.
Importantly, a bent nanotube was used in that experiment, 
allowing for an interpretation of certain features of the magnetotransport  
data, but hindering the direct comparison with our results
corresponding to a straight CNT. 
Nevertheless,  effects from the bend might be unimportant 
when the external magnetic field is perpendicular to 
the plane of the bent CNT, and therefore it makes sense 
to compare our results to the experimental data corresponding
to that case.
(Bend-induced effects will be investigated in future work.)

Figure 3c of Ref. \onlinecite{FeiPei} shows the leakage current as a function of transverse external magnetic field ($B_x$ in our work) 
and (1,1)-(0,2) energy detuning ($\epsilon$ in our work).
The detuning range where our model, neglecting states lying above
the lower-energy doublets, might be relevant 
is approximately the window $[0,1.5]$ eV.
(In our model, this corresponds to 
$-1.5 \, {\rm eV} < \epsilon < 0\, {\rm eV}$.)
The leakage current measured in this range  
clearly shows a resonant peak as a function of detuning at 
$\epsilon \approx 0$, similarly to our result shown in Fig. \ref{fig:detuning}.
However, it is hard to judge whether the predicted hot-spot-type
dependence of the current on the magnetic field strength $B_x$
is present in the
experimental data or not. 
Even if it is, it is certainly blurred by effects not taken into account in 
our model, perhaps by  the interplay of
coherent and inelastic interdot tunneling.

For sufficiently strong negative detuning, the leakage current
due to coherent hybridization
between the (1,1) states and $S_g$ might be overcome by
the leakage current due to
energetically downhill inelastic tunneling processes 
e.g., assisted by phonon emission. 
This latter case is discussed in subsection \ref{subsec:inelastic}.

\subsection{Dependence of the leakage current on interdot tunneling}

The dependence of the leakage current on the amplitude $t$ of coherent 
interdot tunneling has not been investigated in
the experiment of Ref. \onlinecite{FeiPei}. 
Such a study could confirm the relevance of the blockade-lifting
mechanism described in the present work:
Our results indicate that  
the area covered by the current hot spot of Figs. \ref{fig:Setup}c and
\ref{fig:lepedo}
increases, and the position of the hot spot along the $B_x$ axis
is shifted towards larger $B_x$ values, if the 
gate-tunable interdot tunneling matrix element $t$ is increased.

\subsection{Regime of inelastic interdot tunneling} 
\label{subsec:inelastic}

As discussed in subsection \ref{subsec:detuning}, 
at large (1,1)-(0,2) energy detuning $\epsilon \gg t$, 
energetically downhill inelastic (e.g., phonon-emission-mediated)
tunneling processes might dominate the leakage current. 
Jouravlev and Nazarov derived a particularly simple
formula\cite{Jouravlev} for the current in this case, 
expressed as a function of the unit vectors $\vec n_{L}$
and $\vec n_R$
associated to the effective magnetic fields in the two dots:
\bnen
\label{eq:inelastic}
I = \frac{e\Gamma_{\rm in}}{4} \left( \vec n_L \times \vec n_R\right)^2,
\eden
where $\Gamma_{\rm in}$ is the inelastic tunneling rate characterizing the
$S \to S_g$ tunneling process.

\begin{figure*}
\includegraphics[width=0.9\textwidth]{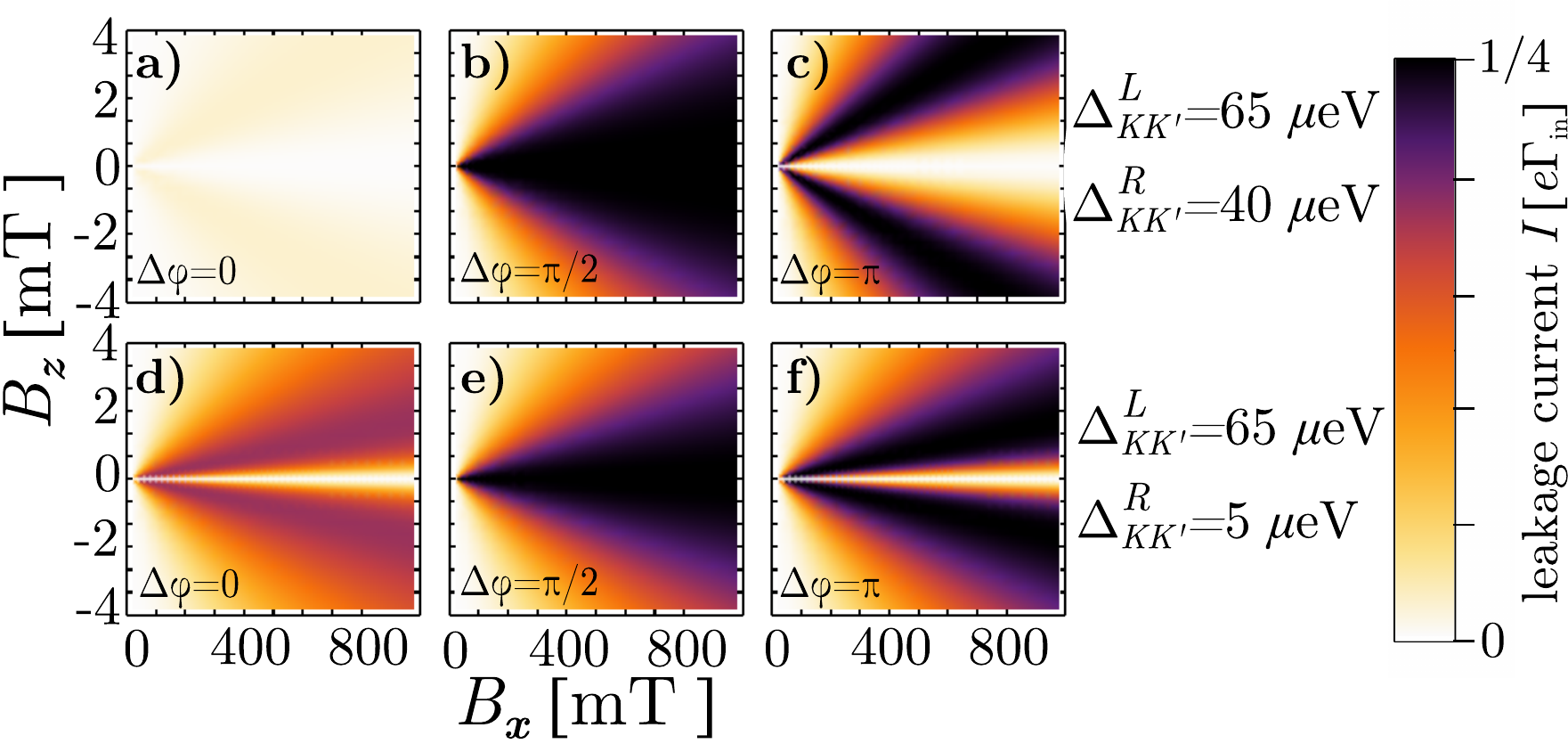}
\caption{\label{fig:inelastic}
(Color online)
The case of inelastic interdot tunneling.
Leakage current as a function of transverse ($B_x$) and
longitudinal ($B_z$) external magnetic field for various
values of the valley-mixing matrix elements
$\Delta_{KK'}^{L}$ and $\Delta_{KK'}^{R}$ in the two dots. 
Each row of graphs is obtained using the valley-mixing
matrix element magnitudes given at the right end of the row. 
The difference $\Delta \varphi$ of the complex phases
of the valley-mixing matrix elements is shown in each graph
at the bottom left corner. 
Parameters: 
$g_s = 2$, $g_v = 54$, $\Delta_{\rm so} =$370$\;\mu$eV.
The plots are obtained by evaluating Eq. \eqref{eq:inelastic}.
}
\end{figure*}

We use this formula to evaluate the leakage current as a function
of longitudal and transverse external magnetic field for different 
values of the valley-mixing matrix elements.
The results are shown in Fig.~\ref{fig:inelastic}.

We note that Eq. \eqref{eq:inelastic} is valid if the magnitudes
of the effective magnetic fields exceed the exchange splitting
within the (1,1) charge configuration, i.e., if
$\beff_L,\beff_R \gg t^2/\epsilon$.


\section{Discussion}
\label{sec:discussion}

\subsection{Role of electron-electron interaction}
\label{subsec:extensions}

Throughout this work we have disregarded 
the (0,2) triplet states, which are typically energetically separated from the
(0,2) ground state $S_g$ by a large exchange gap 
$J_{(0,2)}$\cite{Hanson-rmp}. 
However, two electrons in a CNT QD might form
a Wigner 
molecule\cite{Wunsch,SecchiRontani,Secchi2,vonStecher,Steele-cntdqd,Pecker-wignermolecule}
due to the
strong Coulomb repulsion between electrons and
effective one-dimensional nature of the CNT,
which implies a drastic reduction of the exchange gap $J_{(0,2)}$
in a Pauli-blockaded DQD.
Our description of the current hot spot effect, 
which disregards the (0,2) triplet states, is valid only if the 
hybridization between the (1,1) states and the (0,2) triplet states is negligible,
i.e., if $t \ll J_{(0,2)}$.
This seems to be the case in the spin-valley blockade
experiments of Churchill et al.\cite{ChurchillPRL,ChurchillNPhys}.
The (0,2) exchange gap is very large, comparable to the
fundamental gap of the CNT, in the experiments
reported in Refs. \onlinecite{FeiPei,Laird}, where
n-p type DQDs are used.

Another mechanism not taken into account in our model
is intervalley Coulomb scattering\cite{Ando-valley,Wunsch,SecchiRontani,Weiss,vonStecher,Secchi-intervalley,Pecker-wignermolecule,Cleuziou}, 
arising from the short-range (on-site) contribution of the electron-electron 
interaction.
This mechanism
can mix the (0,2) singlet ground state with higher-lying (0,2) states.
Neglecting this mixing is appropriate as long as the energy scale of 
the corresponding intervalley Coulomb matrix elements is
much smaller than the spin-orbit gap $\Delta_{\rm so}$ separating the 
states in question.

\subsection{Relevance of the results}

The fact that the valley-mixing matrix elements influence the
shape of the current hot spot might be helpful to experimentally identify
the magnitudes and the relative phase of the complex matrix elements
$\Delta_{KK'}^{L}$ and $\Delta_{KK'}^{R}$.
Spatial inhomogeneities of valley-mixing effects play an important
role in schemes proposed recently for electrical manipulation of
single-electron valley- and spin-valley qubits in CNTs\cite{FlensbergMarcus,
Palyi-valley-resonance}.
A spin-valley blockade measurement in the
considered parameter range could be used to explore such 
inhomogeneities.
Furthermore, a difference between the valley-mixing matrix elements 
$\Delta_{KK'}^{L}$ and $\Delta_{KK'}^{R}$ and
the corresponding effective magnetic fields
$\vbeff_L$ and $\vbeff_R$ allows for 
coherent control of  singlet-triplet spin-valley 
qubits,
in a similar fashion as a spatially varying hyperfine or external 
magnetic field allows for singlet-triplet spin qubit 
manipulation\cite{Petta-science}.

Our results are relevant for blockade-based 
experimental applications. 
One example is 
spin-valley qubit initialization and readout\cite{FeiPei,Laird}.
Another example is the dc electronic detection\cite{Ohm} of
the motion of a suspended CNT that acts as a
string-like mechanical resonator, a scheme which is based
on the interaction 
between the spin-valley qubit and the bending phonon modes
\cite{RudnerRashba,Palyi-spinphonon}.
For both applications, it is essential that the leakage current 
is small in the absence of ac driving. 
In this work, we have identified regions in the parameter space
where the leakage current is nonperturbatively large even in
the absence of ac driving;  
qubit initialization/readout and qubit-based nanomechanical motion
detection is possible only outside this parameter regime.

\subsection{Conclusion}

In conclusion, we have shown that valley-mixing,
due to e.g., short-range impurities, can completely lift the 
spin-valley blockade and hence induce a large leakage current 
in carbon nanotube double quantum dots,
if assisted by an appropriately tuned external magnetic field applied
transversally to the tube axis. 
Measurement of the magnetic field dependence of the leakage current could 
provide information about the spatial variation of the valley-mixing
matrix element.
Our study establishes the parameter range 
(magnetic field vector, interdot tunneling, valley-mixing matrix elements)
 where weakly disordered 
CNT DQDs are suited for blockade-based experimental applications
such as qubit initialization/readout and nanomechanical motion
detection. 

\begin{acknowledgments}
We thank J. Danon, 
A. Kiss, E. Laird, and F. Simon for useful discussions. 
Funding from 
the EU Marie Curie Career Integration Grant CIG-293834 (CarbonQubits),
the OTKA Grant PD 100373 and the EU GEOMDISS project 
is acknowledged.
A.~ P.~ is supported by the J\'anos 
Bolyai Research Scholarship of the Hungarian Academy of Sciences.

\end{acknowledgments}


\bibliography{kramersqubit}

\begin{thebibliography}{47}
\expandafter\ifx\csname natexlab\endcsname\relax\def\natexlab#1{#1}\fi
\expandafter\ifx\csname bibnamefont\endcsname\relax
  \def\bibnamefont#1{#1}\fi
\expandafter\ifx\csname bibfnamefont\endcsname\relax
  \def\bibfnamefont#1{#1}\fi
\expandafter\ifx\csname citenamefont\endcsname\relax
  \def\citenamefont#1{#1}\fi
\expandafter\ifx\csname url\endcsname\relax
  \def\url#1{\texttt{#1}}\fi
\expandafter\ifx\csname urlprefix\endcsname\relax\def\urlprefix{URL }\fi
\providecommand{\bibinfo}[2]{#2}
\providecommand{\eprint}[2][]{\url{#2}}

\bibitem[{\citenamefont{Loss and DiVincenzo}(1998)}]{Loss-divincenzo}
\bibinfo{author}{\bibfnamefont{D.}~\bibnamefont{Loss}} \bibnamefont{and}
  \bibinfo{author}{\bibfnamefont{D.~P.} \bibnamefont{DiVincenzo}},
  \bibinfo{journal}{Phys. Rev. A} \textbf{\bibinfo{volume}{57}},
  \bibinfo{pages}{120} (\bibinfo{year}{1998}).

\bibitem[{\citenamefont{Elzerman et~al.}(2004)\citenamefont{Elzerman, Hanson,
  van Beveren, Witkamp, Vandersypen, and Kouwenhoven}}]{Elzerman-nature}
\bibinfo{author}{\bibfnamefont{J.~M.} \bibnamefont{Elzerman}},
  \bibinfo{author}{\bibfnamefont{R.}~\bibnamefont{Hanson}},
  \bibinfo{author}{\bibfnamefont{L.~H.~W.} \bibnamefont{van Beveren}},
  \bibinfo{author}{\bibfnamefont{B.}~\bibnamefont{Witkamp}},
  \bibinfo{author}{\bibfnamefont{L.~M.~K.} \bibnamefont{Vandersypen}},
  \bibnamefont{and} \bibinfo{author}{\bibfnamefont{L.~P.}
  \bibnamefont{Kouwenhoven}}, \bibinfo{journal}{Nature}
  \textbf{\bibinfo{volume}{430}}, \bibinfo{pages}{431} (\bibinfo{year}{2004}).

\bibitem[{\citenamefont{Petta et~al.}(2005)\citenamefont{Petta, Johnson,
  Taylor, Laird, Yacoby, Lukin, Marcus, Hanson, and Gossard}}]{Petta-science}
\bibinfo{author}{\bibfnamefont{J.~R.} \bibnamefont{Petta}},
  \bibinfo{author}{\bibfnamefont{A.~C.} \bibnamefont{Johnson}},
  \bibinfo{author}{\bibfnamefont{J.~M.} \bibnamefont{Taylor}},
  \bibinfo{author}{\bibfnamefont{E.~A.} \bibnamefont{Laird}},
  \bibinfo{author}{\bibfnamefont{A.}~\bibnamefont{Yacoby}},
  \bibinfo{author}{\bibfnamefont{M.~D.} \bibnamefont{Lukin}},
  \bibinfo{author}{\bibfnamefont{C.~M.} \bibnamefont{Marcus}},
  \bibinfo{author}{\bibfnamefont{M.~P.} \bibnamefont{Hanson}},
  \bibnamefont{and} \bibinfo{author}{\bibfnamefont{A.~C.}
  \bibnamefont{Gossard}}, \bibinfo{journal}{Science}
  \textbf{\bibinfo{volume}{309}}, \bibinfo{pages}{2180} (\bibinfo{year}{2005}).

\bibitem[{\citenamefont{Koppens et~al.}(2006)\citenamefont{Koppens, Buizert,
  Tielrooij, Vink, Nowack, Meunier, Kouwenhoven, and
  Vandersypen}}]{Koppens-esr}
\bibinfo{author}{\bibfnamefont{F.~H.~L.} \bibnamefont{Koppens}},
  \bibinfo{author}{\bibfnamefont{C.}~\bibnamefont{Buizert}},
  \bibinfo{author}{\bibfnamefont{K.~J.} \bibnamefont{Tielrooij}},
  \bibinfo{author}{\bibfnamefont{I.~T.} \bibnamefont{Vink}},
  \bibinfo{author}{\bibfnamefont{K.~C.} \bibnamefont{Nowack}},
  \bibinfo{author}{\bibfnamefont{T.}~\bibnamefont{Meunier}},
  \bibinfo{author}{\bibfnamefont{L.~P.} \bibnamefont{Kouwenhoven}},
  \bibnamefont{and} \bibinfo{author}{\bibfnamefont{L.~M.~K.}
  \bibnamefont{Vandersypen}}, \bibinfo{journal}{Nature}
  \textbf{\bibinfo{volume}{442}}, \bibinfo{pages}{766} (\bibinfo{year}{2006}).

\bibitem[{\citenamefont{Nowack et~al.}(2007)\citenamefont{Nowack, Koppens,
  Nazarov, and Vandersypen}}]{Nowack-esr}
\bibinfo{author}{\bibfnamefont{K.~C.} \bibnamefont{Nowack}},
  \bibinfo{author}{\bibfnamefont{F.~H.~L.} \bibnamefont{Koppens}},
  \bibinfo{author}{\bibfnamefont{Y.~V.} \bibnamefont{Nazarov}},
  \bibnamefont{and} \bibinfo{author}{\bibfnamefont{L.~M.~K.}
  \bibnamefont{Vandersypen}}, \bibinfo{journal}{Science}
  \textbf{\bibinfo{volume}{318}}, \bibinfo{pages}{1430} (\bibinfo{year}{2007}).

\bibitem[{\citenamefont{Hanson et~al.}(2007)\citenamefont{Hanson, Kouwenhoven,
  Petta, Tarucha, and Vandersypen}}]{Hanson-rmp}
\bibinfo{author}{\bibfnamefont{R.}~\bibnamefont{Hanson}},
  \bibinfo{author}{\bibfnamefont{L.~P.} \bibnamefont{Kouwenhoven}},
  \bibinfo{author}{\bibfnamefont{J.~R.} \bibnamefont{Petta}},
  \bibinfo{author}{\bibfnamefont{S.}~\bibnamefont{Tarucha}}, \bibnamefont{and}
  \bibinfo{author}{\bibfnamefont{L.~M.~K.} \bibnamefont{Vandersypen}},
  \bibinfo{journal}{Reviews of Modern Physics} \textbf{\bibinfo{volume}{79}},
  \bibinfo{pages}{1217} (\bibinfo{year}{2007}).

\bibitem[{\citenamefont{Ono et~al.}(2002)\citenamefont{Ono, Austing, Tokura,
  and Tarucha}}]{Ono-spinblockade}
\bibinfo{author}{\bibfnamefont{K.}~\bibnamefont{Ono}},
  \bibinfo{author}{\bibfnamefont{D.~G.} \bibnamefont{Austing}},
  \bibinfo{author}{\bibfnamefont{Y.}~\bibnamefont{Tokura}}, \bibnamefont{and}
  \bibinfo{author}{\bibfnamefont{S.}~\bibnamefont{Tarucha}},
  \bibinfo{journal}{Science} \textbf{\bibinfo{volume}{297}},
  \bibinfo{pages}{1313} (\bibinfo{year}{2002}).

\bibitem[{\citenamefont{Koppens et~al.}(2005)\citenamefont{Koppens, Folk,
  Elzerman, Hanson, van Beveren, Vink, Tranitz, Wegscheider, Kouwenhoven, and
  Vandersypen}}]{Koppens-spinblockade}
\bibinfo{author}{\bibfnamefont{F.~H.~L.} \bibnamefont{Koppens}},
  \bibinfo{author}{\bibfnamefont{J.~A.} \bibnamefont{Folk}},
  \bibinfo{author}{\bibfnamefont{J.~M.} \bibnamefont{Elzerman}},
  \bibinfo{author}{\bibfnamefont{R.}~\bibnamefont{Hanson}},
  \bibinfo{author}{\bibfnamefont{L.~H.~W.} \bibnamefont{van Beveren}},
  \bibinfo{author}{\bibfnamefont{T.}~\bibnamefont{Vink}},
  \bibinfo{author}{\bibfnamefont{H.~P.} \bibnamefont{Tranitz}},
  \bibinfo{author}{\bibfnamefont{W.}~\bibnamefont{Wegscheider}},
  \bibinfo{author}{\bibfnamefont{L.~P.} \bibnamefont{Kouwenhoven}},
  \bibnamefont{and} \bibinfo{author}{\bibfnamefont{L.~M.~K.}
  \bibnamefont{Vandersypen}}, \bibinfo{journal}{Science}
  \textbf{\bibinfo{volume}{309}}, \bibinfo{pages}{1346} (\bibinfo{year}{2005}).

\bibitem[{\citenamefont{Nadj-Perge et~al.}(2010)\citenamefont{Nadj-Perge,
  Frolov, van Tilburg, Danon, Nazarov, Algra, Bakkers, and
  Kouwenhoven}}]{NadjPerge-disentangling}
\bibinfo{author}{\bibfnamefont{S.}~\bibnamefont{Nadj-Perge}},
  \bibinfo{author}{\bibfnamefont{S.~M.} \bibnamefont{Frolov}},
  \bibinfo{author}{\bibfnamefont{J.~W.~W.} \bibnamefont{van Tilburg}},
  \bibinfo{author}{\bibfnamefont{J.}~\bibnamefont{Danon}},
  \bibinfo{author}{\bibfnamefont{Y.~V.} \bibnamefont{Nazarov}},
  \bibinfo{author}{\bibfnamefont{R.}~\bibnamefont{Algra}},
  \bibinfo{author}{\bibfnamefont{E.~P. A.~M.} \bibnamefont{Bakkers}},
  \bibnamefont{and} \bibinfo{author}{\bibfnamefont{L.~P.}
  \bibnamefont{Kouwenhoven}}, \bibinfo{journal}{Phys. Rev. B}
  \textbf{\bibinfo{volume}{81}}, \bibinfo{pages}{201305}
  (\bibinfo{year}{2010}).

\bibitem[{\citenamefont{Ohm et~al.}(2012)\citenamefont{Ohm, Stampfer,
  Splettstoesser, and Wegewijs}}]{Ohm}
\bibinfo{author}{\bibfnamefont{C.}~\bibnamefont{Ohm}},
  \bibinfo{author}{\bibfnamefont{C.}~\bibnamefont{Stampfer}},
  \bibinfo{author}{\bibfnamefont{J.}~\bibnamefont{Splettstoesser}},
  \bibnamefont{and} \bibinfo{author}{\bibfnamefont{M.~R.}
  \bibnamefont{Wegewijs}}, \bibinfo{journal}{Appl. Phys. Lett.}
  \textbf{\bibinfo{volume}{100}}, \bibinfo{pages}{143103}
  (\bibinfo{year}{2012}).

\bibitem[{Note1()}]{Note1}
\bibinfo{note}{Throughout this work, we focus on \protect \emph
  {semiconducting} nanotubes, as metallic tubes are not suitable for hosting
  electrostatically defined quantum dots because of the absence of an energy
  gap.}

\bibitem[{\citenamefont{Kuemmeth et~al.}(2008)\citenamefont{Kuemmeth, Ilani,
  Ralph, and McEuen}}]{Kuemmeth}
\bibinfo{author}{\bibfnamefont{F.}~\bibnamefont{Kuemmeth}},
  \bibinfo{author}{\bibfnamefont{S.}~\bibnamefont{Ilani}},
  \bibinfo{author}{\bibfnamefont{D.~C.} \bibnamefont{Ralph}}, \bibnamefont{and}
  \bibinfo{author}{\bibfnamefont{P.~L.} \bibnamefont{McEuen}},
  \bibinfo{journal}{Nature} \textbf{\bibinfo{volume}{452}},
  \bibinfo{pages}{448} (\bibinfo{year}{2008}).

\bibitem[{\citenamefont{Jespersen
  et~al.}(2011{\natexlab{a}})\citenamefont{Jespersen, Grove-Rasmussen, Paaske,
  Muraki, Fujisawa, Nyg�rd, and Flensberg}}]{Jespersen}
\bibinfo{author}{\bibfnamefont{T.~S.} \bibnamefont{Jespersen}},
  \bibinfo{author}{\bibfnamefont{K.}~\bibnamefont{Grove-Rasmussen}},
  \bibinfo{author}{\bibfnamefont{J.}~\bibnamefont{Paaske}},
  \bibinfo{author}{\bibfnamefont{K.}~\bibnamefont{Muraki}},
  \bibinfo{author}{\bibfnamefont{T.}~\bibnamefont{Fujisawa}},
  \bibinfo{author}{\bibfnamefont{J.}~\bibnamefont{Nyg�rd}}, \bibnamefont{and}
  \bibinfo{author}{\bibfnamefont{K.}~\bibnamefont{Flensberg}},
  \bibinfo{journal}{Nat. Phys} \textbf{\bibinfo{volume}{7}},
  \bibinfo{pages}{348} (\bibinfo{year}{2011}{\natexlab{a}}).

\bibitem[{\citenamefont{Steele et~al.}(2013)\citenamefont{Steele, Pei, Laird,
  Jol, Meerwaldt, and Kouwenhoven}}]{Steele-strongsoi}
\bibinfo{author}{\bibfnamefont{G.}~\bibnamefont{Steele}},
  \bibinfo{author}{\bibfnamefont{F.}~\bibnamefont{Pei}},
  \bibinfo{author}{\bibfnamefont{E.}~\bibnamefont{Laird}},
  \bibinfo{author}{\bibfnamefont{J.}~\bibnamefont{Jol}},
  \bibinfo{author}{\bibfnamefont{H.}~\bibnamefont{Meerwaldt}},
  \bibnamefont{and}
  \bibinfo{author}{\bibfnamefont{L.}~\bibnamefont{Kouwenhoven}},
  \bibinfo{journal}{Nat. Comm.} \textbf{\bibinfo{volume}{4}},
  \bibinfo{pages}{1573} (\bibinfo{year}{2013}).

\bibitem[{\citenamefont{Flensberg and Marcus}(2010)}]{FlensbergMarcus}
\bibinfo{author}{\bibfnamefont{K.}~\bibnamefont{Flensberg}} \bibnamefont{and}
  \bibinfo{author}{\bibfnamefont{C.~M.} \bibnamefont{Marcus}},
  \bibinfo{journal}{Phys. Rev. B} \textbf{\bibinfo{volume}{81}},
  \bibinfo{pages}{195418} (\bibinfo{year}{2010}).

\bibitem[{\citenamefont{Laird et~al.}(2013)\citenamefont{Laird, Pei, and
  Kouwenhoven}}]{Laird}
\bibinfo{author}{\bibfnamefont{E.~A.} \bibnamefont{Laird}},
  \bibinfo{author}{\bibfnamefont{F.}~\bibnamefont{Pei}}, \bibnamefont{and}
  \bibinfo{author}{\bibfnamefont{L.~P.} \bibnamefont{Kouwenhoven}},
  \bibinfo{journal}{Nature Nanotechnology} \textbf{\bibinfo{volume}{8}},
  \bibinfo{pages}{565�568} (\bibinfo{year}{2013}).

\bibitem[{\citenamefont{Pei et~al.}(2012)\citenamefont{Pei, Laird, Steele, and
  Kouwenhoven}}]{FeiPei}
\bibinfo{author}{\bibfnamefont{F.}~\bibnamefont{Pei}},
  \bibinfo{author}{\bibfnamefont{E.~A.} \bibnamefont{Laird}},
  \bibinfo{author}{\bibfnamefont{G.~A.} \bibnamefont{Steele}},
  \bibnamefont{and} \bibinfo{author}{\bibfnamefont{L.~P.}
  \bibnamefont{Kouwenhoven}}, \bibinfo{journal}{Nature Nanotechnology}
  \textbf{\bibinfo{volume}{7}}, \bibinfo{pages}{630 } (\bibinfo{year}{2012}).

\bibitem[{\citenamefont{P\'{a}lyi and Burkard}(2009)}]{Palyi-hyperfine}
\bibinfo{author}{\bibfnamefont{A.}~\bibnamefont{P\'{a}lyi}} \bibnamefont{and}
  \bibinfo{author}{\bibfnamefont{G.}~\bibnamefont{Burkard}},
  \bibinfo{journal}{Phys. Rev. B} \textbf{\bibinfo{volume}{80}},
  \bibinfo{pages}{201404} (\bibinfo{year}{2009}).

\bibitem[{\citenamefont{P\'{a}lyi and Burkard}(2010)}]{Palyi-cnt-spinblockade}
\bibinfo{author}{\bibfnamefont{A.}~\bibnamefont{P\'{a}lyi}} \bibnamefont{and}
  \bibinfo{author}{\bibfnamefont{G.}~\bibnamefont{Burkard}},
  \bibinfo{journal}{Phys. Rev. B} \textbf{\bibinfo{volume}{82}},
  \bibinfo{pages}{155424} (\bibinfo{year}{2010}).

\bibitem[{\citenamefont{Buitelaar et~al.}(2008)\citenamefont{Buitelaar,
  Fransson, Cantone, Smith, Anderson, Jones, Ardavan, Khlobystov, Watt,
  Porfyrakis et~al.}}]{Buitelaar}
\bibinfo{author}{\bibfnamefont{M.}~\bibnamefont{Buitelaar}},
  \bibinfo{author}{\bibfnamefont{J.}~\bibnamefont{Fransson}},
  \bibinfo{author}{\bibfnamefont{A.}~\bibnamefont{Cantone}},
  \bibinfo{author}{\bibfnamefont{C.}~\bibnamefont{Smith}},
  \bibinfo{author}{\bibfnamefont{D.}~\bibnamefont{Anderson}},
  \bibinfo{author}{\bibfnamefont{G.}~\bibnamefont{Jones}},
  \bibinfo{author}{\bibfnamefont{A.}~\bibnamefont{Ardavan}},
  \bibinfo{author}{\bibfnamefont{A.}~\bibnamefont{Khlobystov}},
  \bibinfo{author}{\bibfnamefont{A.}~\bibnamefont{Watt}},
  \bibinfo{author}{\bibfnamefont{K.}~\bibnamefont{Porfyrakis}},
  \bibnamefont{et~al.}, \bibinfo{journal}{Phys. Rev. B}
  \textbf{\bibinfo{volume}{77}}, \bibinfo{pages}{245439}
  (\bibinfo{year}{2008}).

\bibitem[{\citenamefont{Churchill
  et~al.}(2009{\natexlab{a}})\citenamefont{Churchill, Kuemmeth, Harlow,
  Bestwick, Rashba, Flensberg, Stwertka, Taychatanapat, Watson, and
  Marcus}}]{ChurchillPRL}
\bibinfo{author}{\bibfnamefont{H.~O.~H.} \bibnamefont{Churchill}},
  \bibinfo{author}{\bibfnamefont{F.}~\bibnamefont{Kuemmeth}},
  \bibinfo{author}{\bibfnamefont{J.~W.} \bibnamefont{Harlow}},
  \bibinfo{author}{\bibfnamefont{A.~J.} \bibnamefont{Bestwick}},
  \bibinfo{author}{\bibfnamefont{E.~I.} \bibnamefont{Rashba}},
  \bibinfo{author}{\bibfnamefont{K.}~\bibnamefont{Flensberg}},
  \bibinfo{author}{\bibfnamefont{C.~H.} \bibnamefont{Stwertka}},
  \bibinfo{author}{\bibfnamefont{T.}~\bibnamefont{Taychatanapat}},
  \bibinfo{author}{\bibfnamefont{S.~K.} \bibnamefont{Watson}},
  \bibnamefont{and} \bibinfo{author}{\bibfnamefont{C.~M.}
  \bibnamefont{Marcus}}, \bibinfo{journal}{Phys. Rev. Lett.}
  \textbf{\bibinfo{volume}{102}}, \bibinfo{pages}{166802}
  (\bibinfo{year}{2009}{\natexlab{a}}).

\bibitem[{\citenamefont{Churchill
  et~al.}(2009{\natexlab{b}})\citenamefont{Churchill, Bestwick, Harlow,
  Kuemmeth, Marcos, Stwertka, Watson, and Marcus}}]{ChurchillNPhys}
\bibinfo{author}{\bibfnamefont{H.~O.~H.} \bibnamefont{Churchill}},
  \bibinfo{author}{\bibfnamefont{A.~J.} \bibnamefont{Bestwick}},
  \bibinfo{author}{\bibfnamefont{J.~W.} \bibnamefont{Harlow}},
  \bibinfo{author}{\bibfnamefont{F.}~\bibnamefont{Kuemmeth}},
  \bibinfo{author}{\bibfnamefont{D.}~\bibnamefont{Marcos}},
  \bibinfo{author}{\bibfnamefont{C.~H.} \bibnamefont{Stwertka}},
  \bibinfo{author}{\bibfnamefont{S.~K.} \bibnamefont{Watson}},
  \bibnamefont{and} \bibinfo{author}{\bibfnamefont{C.~M.}
  \bibnamefont{Marcus}}, \bibinfo{journal}{Nature Physics}
  \textbf{\bibinfo{volume}{5}}, \bibinfo{pages}{321}
  (\bibinfo{year}{2009}{\natexlab{b}}).

\bibitem[{\citenamefont{Chorley et~al.}(2011)\citenamefont{Chorley, Giavaras,
  Wabnig, Jones, Smith, Briggs, and Buitelaar}}]{Chorley}
\bibinfo{author}{\bibfnamefont{S.~J.} \bibnamefont{Chorley}},
  \bibinfo{author}{\bibfnamefont{G.}~\bibnamefont{Giavaras}},
  \bibinfo{author}{\bibfnamefont{J.}~\bibnamefont{Wabnig}},
  \bibinfo{author}{\bibfnamefont{G.~A.~C.} \bibnamefont{Jones}},
  \bibinfo{author}{\bibfnamefont{C.~G.} \bibnamefont{Smith}},
  \bibinfo{author}{\bibfnamefont{G.~A.~D.} \bibnamefont{Briggs}},
  \bibnamefont{and} \bibinfo{author}{\bibfnamefont{M.~R.}
  \bibnamefont{Buitelaar}}, \bibinfo{journal}{Phys. Rev. Lett.}
  \textbf{\bibinfo{volume}{106}}, \bibinfo{pages}{206801}
  (\bibinfo{year}{2011}).

\bibitem[{\citenamefont{Jouravlev and Nazarov}(2006)}]{Jouravlev}
\bibinfo{author}{\bibfnamefont{O.~N.} \bibnamefont{Jouravlev}}
  \bibnamefont{and} \bibinfo{author}{\bibfnamefont{Y.~V.}
  \bibnamefont{Nazarov}}, \bibinfo{journal}{Phys. Rev. Lett.}
  \textbf{\bibinfo{volume}{96}}, \bibinfo{pages}{176804}
  (\bibinfo{year}{2006}).

\bibitem[{\citenamefont{Coish and Qassemi}(2011)}]{Coish}
\bibinfo{author}{\bibfnamefont{W.~A.} \bibnamefont{Coish}} \bibnamefont{and}
  \bibinfo{author}{\bibfnamefont{F.}~\bibnamefont{Qassemi}},
  \bibinfo{journal}{Phys. Rev. B} \textbf{\bibinfo{volume}{84}},
  \bibinfo{pages}{245407} (\bibinfo{year}{2011}).

\bibitem[{\citenamefont{Rudner and Rashba}(2010)}]{RudnerRashba}
\bibinfo{author}{\bibfnamefont{M.~S.} \bibnamefont{Rudner}} \bibnamefont{and}
  \bibinfo{author}{\bibfnamefont{E.~I.} \bibnamefont{Rashba}},
  \bibinfo{journal}{Phys. Rev. B} \textbf{\bibinfo{volume}{81}},
  \bibinfo{pages}{125426} (\bibinfo{year}{2010}).

\bibitem[{\citenamefont{P\'{a}lyi et~al.}(2012)\citenamefont{P\'{a}lyi, Struck,
  Rudner, Flensberg, and Burkard}}]{Palyi-spinphonon}
\bibinfo{author}{\bibfnamefont{A.}~\bibnamefont{P\'{a}lyi}},
  \bibinfo{author}{\bibfnamefont{P.~R.} \bibnamefont{Struck}},
  \bibinfo{author}{\bibfnamefont{M.}~\bibnamefont{Rudner}},
  \bibinfo{author}{\bibfnamefont{K.}~\bibnamefont{Flensberg}},
  \bibnamefont{and} \bibinfo{author}{\bibfnamefont{G.}~\bibnamefont{Burkard}},
  \bibinfo{journal}{Phys. Rev. Lett.} \textbf{\bibinfo{volume}{108}},
  \bibinfo{pages}{206811} (\bibinfo{year}{2012}).

\bibitem[{\citenamefont{Reynoso and Flensberg}(2011)}]{Reynoso1}
\bibinfo{author}{\bibfnamefont{A.~A.} \bibnamefont{Reynoso}} \bibnamefont{and}
  \bibinfo{author}{\bibfnamefont{K.}~\bibnamefont{Flensberg}},
  \bibinfo{journal}{Phys. Rev. B} \textbf{\bibinfo{volume}{84}},
  \bibinfo{pages}{205449} (\bibinfo{year}{2011}).

\bibitem[{\citenamefont{Reynoso and Flensberg}(2012)}]{Reynoso2}
\bibinfo{author}{\bibfnamefont{A.~A.} \bibnamefont{Reynoso}} \bibnamefont{and}
  \bibinfo{author}{\bibfnamefont{K.}~\bibnamefont{Flensberg}},
  \bibinfo{journal}{Phys. Rev. B} \textbf{\bibinfo{volume}{85}},
  \bibinfo{pages}{195441} (\bibinfo{year}{2012}).

\bibitem[{\citenamefont{Wunsch}(2009)}]{Wunsch}
\bibinfo{author}{\bibfnamefont{B.}~\bibnamefont{Wunsch}},
  \bibinfo{journal}{Phys. Rev. B} \textbf{\bibinfo{volume}{79}},
  \bibinfo{pages}{235408} (\bibinfo{year}{2009}).

\bibitem[{\citenamefont{Secchi and Rontani}(2009)}]{SecchiRontani}
\bibinfo{author}{\bibfnamefont{A.}~\bibnamefont{Secchi}} \bibnamefont{and}
  \bibinfo{author}{\bibfnamefont{M.}~\bibnamefont{Rontani}},
  \bibinfo{journal}{Phys. Rev. B} \textbf{\bibinfo{volume}{80}},
  \bibinfo{pages}{041404} (\bibinfo{year}{2009}).

\bibitem[{\citenamefont{Weiss et~al.}(2010)\citenamefont{Weiss, Rashba,
  Kuemmeth, Churchill, and Flensberg}}]{Weiss}
\bibinfo{author}{\bibfnamefont{S.}~\bibnamefont{Weiss}},
  \bibinfo{author}{\bibfnamefont{E.~I.} \bibnamefont{Rashba}},
  \bibinfo{author}{\bibfnamefont{F.}~\bibnamefont{Kuemmeth}},
  \bibinfo{author}{\bibfnamefont{H.~O.~H.} \bibnamefont{Churchill}},
  \bibnamefont{and}
  \bibinfo{author}{\bibfnamefont{K.}~\bibnamefont{Flensberg}},
  \bibinfo{journal}{Phys. Rev. B} \textbf{\bibinfo{volume}{82}},
  \bibinfo{pages}{165427} (\bibinfo{year}{2010}).

\bibitem[{\citenamefont{von Stecher et~al.}(2010)\citenamefont{von Stecher,
  Wunsch, Lukin, Demler, and Rey}}]{vonStecher}
\bibinfo{author}{\bibfnamefont{J.}~\bibnamefont{von Stecher}},
  \bibinfo{author}{\bibfnamefont{B.}~\bibnamefont{Wunsch}},
  \bibinfo{author}{\bibfnamefont{M.}~\bibnamefont{Lukin}},
  \bibinfo{author}{\bibfnamefont{E.}~\bibnamefont{Demler}}, \bibnamefont{and}
  \bibinfo{author}{\bibfnamefont{A.~M.} \bibnamefont{Rey}},
  \bibinfo{journal}{Phys. Rev. B} \textbf{\bibinfo{volume}{82}},
  \bibinfo{pages}{125437} (\bibinfo{year}{2010}).

\bibitem[{\citenamefont{Kiss et~al.}(2011)\citenamefont{Kiss, P\'{a}lyi, Ihara,
  Wzietek, Alloul, Simon, Z\'{o}lyomi, Koltai, K\"{u}rti, D\'{o}ra
  et~al.}}]{Kiss}
\bibinfo{author}{\bibfnamefont{A.}~\bibnamefont{Kiss}},
  \bibinfo{author}{\bibfnamefont{A.}~\bibnamefont{P\'{a}lyi}},
  \bibinfo{author}{\bibfnamefont{Y.}~\bibnamefont{Ihara}},
  \bibinfo{author}{\bibfnamefont{P.}~\bibnamefont{Wzietek}},
  \bibinfo{author}{\bibfnamefont{H.}~\bibnamefont{Alloul}},
  \bibinfo{author}{\bibfnamefont{P.}~\bibnamefont{Simon}},
  \bibinfo{author}{\bibfnamefont{V.}~\bibnamefont{Z\'{o}lyomi}},
  \bibinfo{author}{\bibfnamefont{J.}~\bibnamefont{Koltai}},
  \bibinfo{author}{\bibfnamefont{J.}~\bibnamefont{K\"{u}rti}},
  \bibinfo{author}{\bibfnamefont{B.}~\bibnamefont{D\'{o}ra}},
  \bibnamefont{et~al.}, \bibinfo{journal}{Phys. Rev. Lett.}
  \textbf{\bibinfo{volume}{107}}, \bibinfo{pages}{187204}
  (\bibinfo{year}{2011}).

\bibitem[{\citenamefont{P\'alyi and Burkard}(2011)}]{Palyi-valley-resonance}
\bibinfo{author}{\bibfnamefont{A.}~\bibnamefont{P\'alyi}} \bibnamefont{and}
  \bibinfo{author}{\bibfnamefont{G.}~\bibnamefont{Burkard}},
  \bibinfo{journal}{Phys. Rev. Lett.} \textbf{\bibinfo{volume}{106}},
  \bibinfo{pages}{086801} (\bibinfo{year}{2011}).

\bibitem[{\citenamefont{Minot et~al.}(2004)\citenamefont{Minot, Yaish,
  Sazonova, and McEuen}}]{Minot}
\bibinfo{author}{\bibfnamefont{E.~D.} \bibnamefont{Minot}},
  \bibinfo{author}{\bibfnamefont{Y.}~\bibnamefont{Yaish}},
  \bibinfo{author}{\bibfnamefont{V.}~\bibnamefont{Sazonova}}, \bibnamefont{and}
  \bibinfo{author}{\bibfnamefont{P.~L.} \bibnamefont{McEuen}},
  \bibinfo{journal}{Nature} \textbf{\bibinfo{volume}{428}},
  \bibinfo{pages}{536} (\bibinfo{year}{2004}).

\bibitem[{\citenamefont{Jespersen
  et~al.}(2011{\natexlab{b}})\citenamefont{Jespersen, Grove-Rasmussen,
  Flensberg, Paaske, Muraki, Fujisawa, and Nyg\aa{}rd}}]{Jespersen-orbital}
\bibinfo{author}{\bibfnamefont{T.~S.} \bibnamefont{Jespersen}},
  \bibinfo{author}{\bibfnamefont{K.}~\bibnamefont{Grove-Rasmussen}},
  \bibinfo{author}{\bibfnamefont{K.}~\bibnamefont{Flensberg}},
  \bibinfo{author}{\bibfnamefont{J.}~\bibnamefont{Paaske}},
  \bibinfo{author}{\bibfnamefont{K.}~\bibnamefont{Muraki}},
  \bibinfo{author}{\bibfnamefont{T.}~\bibnamefont{Fujisawa}}, \bibnamefont{and}
  \bibinfo{author}{\bibfnamefont{J.}~\bibnamefont{Nyg\aa{}rd}},
  \bibinfo{journal}{Phys. Rev. Lett.} \textbf{\bibinfo{volume}{107}},
  \bibinfo{pages}{186802} (\bibinfo{year}{2011}{\natexlab{b}}).

\bibitem[{\citenamefont{Winkler}(2003)}]{Winkler}
\bibinfo{author}{\bibfnamefont{R.}~\bibnamefont{Winkler}},
  \emph{\bibinfo{title}{Spin-Orbit Coupling Effects in Two-Dimensional Electron
  and Hole Systems}} (\bibinfo{publisher}{Springer-Verlag},
  \bibinfo{year}{2003}).

\bibitem[{\citenamefont{Culcer et~al.}(2012)\citenamefont{Culcer, Saraiva,
  Koiller, Hu, and {Das Sarma}}}]{CulcerPRL}
\bibinfo{author}{\bibfnamefont{D.}~\bibnamefont{Culcer}},
  \bibinfo{author}{\bibfnamefont{A.~L.} \bibnamefont{Saraiva}},
  \bibinfo{author}{\bibfnamefont{B.}~\bibnamefont{Koiller}},
  \bibinfo{author}{\bibfnamefont{X.}~\bibnamefont{Hu}}, \bibnamefont{and}
  \bibinfo{author}{\bibfnamefont{S.}~\bibnamefont{{Das Sarma}}},
  \bibinfo{journal}{Phys. Rev. Lett.} \textbf{\bibinfo{volume}{108}},
  \bibinfo{pages}{126804} (\bibinfo{year}{2012}).

\bibitem[{\citenamefont{Wu and Culcer}(2012)}]{YueWu}
\bibinfo{author}{\bibfnamefont{Y.}~\bibnamefont{Wu}} \bibnamefont{and}
  \bibinfo{author}{\bibfnamefont{D.}~\bibnamefont{Culcer}},
  \bibinfo{journal}{Phys. Rev. B} \textbf{\bibinfo{volume}{86}},
  \bibinfo{pages}{035321} (\bibinfo{year}{2012}).

\bibitem[{\citenamefont{Danon et~al.}(2013)\citenamefont{Danon, Wang, and
  Manchon}}]{Jeroen}
\bibinfo{author}{\bibfnamefont{J.}~\bibnamefont{Danon}},
  \bibinfo{author}{\bibfnamefont{X.}~\bibnamefont{Wang}}, \bibnamefont{and}
  \bibinfo{author}{\bibfnamefont{A.}~\bibnamefont{Manchon}},
  \bibinfo{journal}{Phys. Rev. Lett.} \textbf{\bibinfo{volume}{111}},
  \bibinfo{pages}{066802} (\bibinfo{year}{2013}).

\bibitem[{\citenamefont{Secchi and Rontani}(2010)}]{Secchi2}
\bibinfo{author}{\bibfnamefont{A.}~\bibnamefont{Secchi}} \bibnamefont{and}
  \bibinfo{author}{\bibfnamefont{M.}~\bibnamefont{Rontani}},
  \bibinfo{journal}{Phys. Rev. B} \textbf{\bibinfo{volume}{82}},
  \bibinfo{pages}{035417} (\bibinfo{year}{2010}).

\bibitem[{\citenamefont{Steele et~al.}(2009)\citenamefont{Steele, Gotz, and
  Kouwenhoven}}]{Steele-cntdqd}
\bibinfo{author}{\bibfnamefont{G.}~\bibnamefont{Steele}},
  \bibinfo{author}{\bibfnamefont{G.}~\bibnamefont{Gotz}}, \bibnamefont{and}
  \bibinfo{author}{\bibfnamefont{L.~P.} \bibnamefont{Kouwenhoven}},
  \bibinfo{journal}{Nat. Nanotech.} \textbf{\bibinfo{volume}{4}},
  \bibinfo{pages}{363} (\bibinfo{year}{2009}).

\bibitem[{\citenamefont{Pecker et~al.}(2013)\citenamefont{Pecker, Kuemmeth,
  Secchi, Rontani, Ralph, McEuen, and Ilani}}]{Pecker-wignermolecule}
\bibinfo{author}{\bibfnamefont{S.}~\bibnamefont{Pecker}},
  \bibinfo{author}{\bibfnamefont{F.}~\bibnamefont{Kuemmeth}},
  \bibinfo{author}{\bibfnamefont{A.}~\bibnamefont{Secchi}},
  \bibinfo{author}{\bibfnamefont{M.}~\bibnamefont{Rontani}},
  \bibinfo{author}{\bibfnamefont{D.~C.} \bibnamefont{Ralph}},
  \bibinfo{author}{\bibfnamefont{P.~L.} \bibnamefont{McEuen}},
  \bibnamefont{and} \bibinfo{author}{\bibfnamefont{S.}~\bibnamefont{Ilani}},
  \bibinfo{journal}{Nat. Phys.} \textbf{\bibinfo{volume}{9}}
  (\bibinfo{year}{2013}).

\bibitem[{\citenamefont{Ando}(2006)}]{Ando-valley}
\bibinfo{author}{\bibfnamefont{T.}~\bibnamefont{Ando}},
  \bibinfo{journal}{Journal of the Physical Society of Japan}
  \textbf{\bibinfo{volume}{75}}, \bibinfo{pages}{024707}
  (\bibinfo{year}{2006}).

\bibitem[{\citenamefont{Secchi and Rontani}(2013)}]{Secchi-intervalley}
\bibinfo{author}{\bibfnamefont{A.}~\bibnamefont{Secchi}} \bibnamefont{and}
  \bibinfo{author}{\bibfnamefont{M.}~\bibnamefont{Rontani}},
  \bibinfo{journal}{Phys. Rev. B} \textbf{\bibinfo{volume}{88}},
  \bibinfo{pages}{125403} (\bibinfo{year}{2013}).

\bibitem[{\citenamefont{Cleuziou et~al.}(2013)\citenamefont{Cleuziou, N'Guyen,
  Florens, and Wernsdorfer}}]{Cleuziou}
\bibinfo{author}{\bibfnamefont{J.~P.} \bibnamefont{Cleuziou}},
  \bibinfo{author}{\bibfnamefont{N.~V.} \bibnamefont{N'Guyen}},
  \bibinfo{author}{\bibfnamefont{S.}~\bibnamefont{Florens}}, \bibnamefont{and}
  \bibinfo{author}{\bibfnamefont{W.}~\bibnamefont{Wernsdorfer}},
  \bibinfo{journal}{Phys. Rev. Lett.} \textbf{\bibinfo{volume}{111}},
  \bibinfo{pages}{136803} (\bibinfo{year}{2013}).

\end{thebibliography}

\end{document}